**Strongly Correlated Materials**

By *Emilia Morosan, Douglas Natelson\*, Andriy H. Nevidomskyy* and *Qimiao Si*

Prof. E. Morosan, Prof. D. Natelson, Prof. A. H. Nevidomskyy, Prof. Q. Si
Rice University
Department of Physics and Astronomy MS 61
6100 Main St.
Houston, TX, 77005 (USA)
E-mail: natelson@rice.edu

Strongly correlated materials are profoundly affected by the repulsive electron-electron interaction. This stands in contrast to many commonly used materials such as silicon and aluminum, whose properties are comparatively unaffected by the Coulomb repulsion. 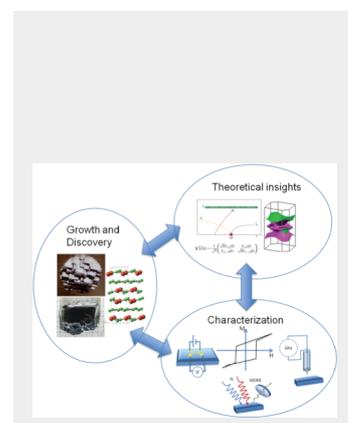 Correlated materials often have remarkable properties and transitions between distinct, competing phases with dramatically different electronic and magnetic orders. These rich phenomena are fascinating from the basic science perspective and offer possibilities for technological applications. This article looks at these materials through the lens of research performed at Rice University. Topics examined include: Quantum phase transitions and quantum criticality in "heavy fermion" materials and the iron pnictide high temperature superconductors; computational *ab initio* methods to examine strongly correlated materials and their interface with analytical theory techniques; layered dichalcogenides as example correlated materials with rich phases (charge density waves, superconductivity, hard ferromagnetism) that may be tuned by composition, pressure, and magnetic field; and nanostructure methods applied to the correlated oxides $VO_2$ and $Fe_3O_4$, where metal-insulator transitions can be manipulated by doping at the nanoscale or driving the system out of equilibrium. We conclude with a discussion of the exciting prospects for this class of materials.

# 1. Introduction

Many of the materials that shape our world and its technologies have properties that are comparatively insensitive to the repulsive interactions between electrons. The electrical, mechanical, and thermal properties of silicon, aluminum, diamond, and even more exotic newcomers to the materials lexicon like graphene, may all be understood reasonably well from the point of view of non-interacting electrons. Because of the Pauli exclusion principle and the delocalized character of their electronic states, built primarily from s and p orbitals, the kinetic energy of the electrons dominates rather than the electron-electron interaction. Even for materials whose defining properties come about through electron-electron interactions, such as itinerant ferromagnets in the magnetoresistive structures that are so prominent in present-day information technology, the interaction effects are often treated as an afterthought. The results are the familiar metals, semiconductors, band insulators, and semimetals of band theory, the building blocks and constituents of the large majority of material systems discussed in this issue.

However, there are many materials in which electron-electron interactions play a centrol role in determining the electronic, magnetic, optical, and sometimes even mechanical properties. This is presaged even at the atomic level in the form of Hund's Rules in the transition metals, when the Coulomb energetic savings makes it worthwhile for d electrons to align their spins, and thus reside in separate spatial orbitals. These and related exchange interactions, inherently based on the combination of the electron-electron repulsion and quantum mechanics, are at the root of magnetic order, which has proven to be of immense technological relevance and deep physical import.

In general, the complex interplay between electron-electron interactions, lattice structure, kinetic energy (via the electron's ability to quantum-mechanically tunnel around in a crystal), and magnetic degrees of freedom is incredibly rich. The result is a competition between distinct ground states, with different symmetries and low energy excitations; and further competition between "local" quantum degrees of freedom associated with individual lattice sites, and longer scale fluctuations and excitations. New phases often appear near "quantum phase transitions", when tuning a parameter such as electronic population or magnetic field tips the balance between one energy scale and one of its competitors. These phases can have surprising and useful properties, such as high temperature superconductivity. Likewise, the transitions between such phases can have technologically relevant features, such as orders-of-magnitude changes in electrical conductivity ("metal-insulator transitions"), perhaps tuned by magnetic field ("colossal magnetoresistance"). At these cusps in behavior, strong correlations can act as a sort of lever arm, through which small changes in controllable parameters can have dramatic consequences on material properties.

The strong electron correlations that are responsible for these rich electronic material properties also make them difficult to study. In the case of relatively weak electron-electron repulsion (compared to the kinetic energy), we have a reasonably good understanding of the electronic behavior based on well-developed perturbative methods such as (k.p) expansion, the local density and Hartree-Fock approximations, Thomas-Fermi screening in metals, etc. In contrast, when the interactions become sizable, as is often the case in the d- and f - electron systems, the portfolio of controlled theoretical methods shrinks dramatically. With the exception of one-dimensional systems for which some very specialized analytical and computational techniques have been developed, the problem remains largely open. Nevertheless, two concepts in particular come to the rescue: Landau's adiabaticity principle, and the paradigm of the so-called "renormalization group".

The Landau principle states that as the strength of the interactions increases adiabatically, the role of the electrons as low energy excitations gets played by effective quasi-particles, with essentially the same qualitative properties but different detailed characteristics. For example, these quasi-particles can acquire a large mass due to their interactions with the surrounding medium. The so-called heavy fermion materials are an extreme example, in which the effective mass of the quasi-particles can reach hundreds of times the bare electron mass. The second concept, renormalizability, originates in high-energy physics. It provides an effective tool for deducing the effective low-temperature behaviour of the interacting system by systematically getting rid of ("integrating out") the high-energy degrees of freedom. This procedure, known as the renormalization group, has proven to be indispensable, in particular in the theory of the phase transitions. Taken together, these two principles provide foundation for how we understand the complex interacting electron systems. In addition, various sophisticated computational techniques have been devised to approximately treat the effect of electron correlations, for example quantum Monte Carlo, functional renormalization and various quantum cluster methods. Combined with various ab initio band theory calculations, this arsenal of theoretical methods allows one not only calculate but also predict materials properties, paving the way towards predictive materials design.

Together with the fast developing experimental and material growth techniques, these theoretical tools herald the approaching possibility of being able to understand, predict, engineer, and control materials with desired electronic properties and tunable transitions between various phases. Building on an ever-increasing foundation, materials synthesis and growth can create materials never before realized. Characterization of these new materials, both in the bulk and through nanoscale methods and techniques, extends our understanding. As pointed out by the national call for a Materials Genome Initiative,[1] we are on the cusp of "materials by design" through the feedback loop of materials theory, computation, discovery/growth, and characterization.

In this review, we will discuss several examples of strongly correlated materials and transitions between their phases that have been examined at Rice and elsewhere. These materials illustrate the diversity in the materials families with varying degrees of electron correlations, as summarized in **Fig. 1**. We begin with a discussion of *quantum criticality and superconductivity at the border of magnetism*, where local degrees of freedom such as magnetic moments can take on added importance due to interactions. The multiplicity in the interactions that involve such microscopic degrees of freedom leads to competing phases. A continuous transformation between such phases at zero temperature gives rise to a quantum critical point, which profoundly influences physical properties at nonzero temperatures and can even nucleate novel phases such as unconventional superconductivity. We then turn to *electronic structure in the presence of correlations*, looking specifically at computational approaches as applied to these complex systems, bridging the gap between analytical theory and predictive materials science modeling. We examine *transition metal dichalcogenides* as a versatile, fascinating family of correlated materials exhibiting diverse, competing ordered states including charge density waves, superconductivity, and hard ferromagnetism. This material family is illustrative of the power and techniques of bulk correlated materials growth. Lastly, we look at *nanostructure techniques* as applied to strongly correlated materials, particularly $VO_2$ and $Fe_3O_4$. The experiments discussed demonstrate how nanoscale methods can permit studies not feasible in macroscopic systems. Equilibrium measurements can give insight into the competition between ordered states, while nanostructures enable nonequilibrium studies that use perturbations to tip the balance between competing phases. We conclude with perspectives and opportunities in this exciting class of materials.

## 2. Quantum Criticality and Superconductivity at the Border of Magnetism

It is readily recognized that the dipole-dipole interaction between electron spins is far too weak ($\lesssim 1$ K) to explain the observed magnetic transition in ferro- and antiferromagnets. The concepts of quantum mechanics, such as the Pauli exclusion principle and the exchange coupling due to Coulomb repulsion between electrons, are necessary to explain these phenomena, which are collectively termed "quantum magnetism". For our purposes, this provides an ideal setting for illustrating the richness of quantum phases in electronic materials. As we shall see below, quantum magnetism is closely tied with strong electronic correlations and emerges as the low-energy physics of many correlated metals and insulators.

Quantum magnetism is perhaps the simplest example of an emergent quantum phase, defined as a non-trivial quantum state that spontaneously emerges at low energies and low temperatures, while the higher-energy degrees of freedom may have very different characteristics. In this example, the effects of electron spins may not be noticeable at higher temperatures, where the dominant physics is dictated by electron charge, but the situation changes radically as the temperature is lowered. The "emergent order" can be viewed as a way for the quantum system to get rid of its entropy at low temperatures, to satisfy the third law of thermodynamics. Sometimes, several possible quantum states compete to get the upper hand at low temperatures, and this leads to a particularly powerful concept of a zero-temperature phase transition between the two states, the so-called quantum phase transition. Just as classical phase transitions (such as the liquid-gas transition) occur due to thermal fluctuations as the temperature is raised, quantum transitions are realized when a non-thermal control parameter, such as pressure or chemical doping, is varied at a very low temperature. A variety of magnetically ordered and paramagnetic phases are being actively studied by physicists, both experimentally and in theoretical models.

To illustrate the competition between different ordered quantum phases, consider the case of antiferromagnetism when the Néel temperature is continuously suppressed to zero by a non-thermal parameter, *e.g.* applied pressure. The threshold pressure $p_c$ defines a quantum critical point (QCP), separating antiferromagnetic and paramagnetic ground states.[6-9] What is particularly intriguing, is that precisely at the QCP, the ground state and low-energy excitations are distinct from those on either side of $p_c$. Quantum criticality thus refers to the low-energy fluctuations at the QCP, which are often composed of emergent collective excitations that are different from those away from $p_c$. The collectivity is characterized by the maximization of entropy.[10, 11] In other words, a quantum critical system is particularly "soft" right at the point of the transition, and is often prone to the formation of new electronic phases.

One example of such a truly emergent phase is superconductivity that forms at the border of antiferromagnetism. **Figure 2a** illustrates the temperature-pressure ($T$-$p$) phase diagram of the heavy-fermion material, $CePd_2Si_2$.[2] The Néel transition temperature is gradually reduced by the external pressure and as it approaches zero, a dome of superconductivity appears in the phase diagram. This behaviour turns out to be ubiquitous and is also observed in the recently discovered iron-based and cuprate high-temperature superconductors as a function of chemical doping, and in the organic charge-transfer salts as a function of pressure. These are illustrated in Figs. 2b-d.

Superconductivity, characterized by zero resistivity and Meissner effect, reflects the condensation of electron pairs.[12] Just like the description of conventional superconductors requires an account of the electronic and phononic excitations in the normal state above the superconducting transition temperature, $T_c$,[13] so, too, the understanding of unconventional superconductivity here necessitates the description of the quantum criticality in the normal state.

From the perspective of correlated electrons, quantum magnetism reflects the repulsive Coulomb interactions between the electrons. Therefore, the emergent unconventional superconductivity at the border of antiferromagnetism is intimately connected to the question of how electron pairing can originate from repulsive electron-electron interactions. Below, we shall first discuss different types of magnetic ordering in strongly correlated electron systems, followed by more detailed account of two classes of correlated materials in particular: the heavy fermion compounds and the iron-based superconductors.

**2.1. Quantum magnetism and correlated electrons**

Magnetism in correlated electron systems comes in several broad categories. The commonly considered limits are specified in terms of the ratio of Coulomb repulsion to kinetic energy, $U/D$. Here $U$ describes the effective on-site Coulomb repulsion, which includes the Hund's coupling if there is more than one relevant orbital, and $D$ is the effective width of the pertinent electronic bands.

When $U/D$ is small, in other words when the Coulomb repulsion is weak compared to the kinetic energy, one can use the non-interacting limit as a reference point and treat Coulomb interaction as a perturbation. This approach gives rise to so-called itinerant magnetism, and the corresponding antiferromagnetic order is often referred to as a spin-density-wave (SDW) order. The canonical example is Cr,[14, 15] which is a good metal and undergoes an SDW transition at 311 K, below which the spins order in a spiral fashion.

When $U/D$ is large, on the other hand, electrons incur a potential energy penalty for hopping between different lattice sites, and for sufficiently large $U$, they become localized, with an interaction-induced gap in the spectrum of electronic excitations. The resulting insulating state is called a Mott insulator, after its discoverer Sir Nevill Mott, who was awarded a Nobel prize for his contribution into our present understanding of strongly correlated systems. In the Mott insulating state, the charge degrees of freedom are frozen, and the accompanying spin degrees of freedom are referred to as local moments, as opposed to the previously discussed itinerant moments. These local moments are coupled with each other through exchange interactions and often take on an antiferromagnetic order. A well-known example is NiO, which is an insulator with a charge gap over 4 eV, even though it would have been metallic in the absence of electron-electron interactions, and has a Néel temperature of over 500 K (Refs. [16-18]).

The third category of quantum magnets lies in between these two extremes, occurring in metallic systems whose electrons are at the boundary between delocalization and localization. Such materials have emerged as fertile hosts for quantum criticality and emergent quantum phenomena, such as unconventional superconductivity. One example is the class of heavy fermion metals, so called because strong electron correlations lead to a hundredfold increase in the mass of charge carriers (fermions) compared to the bare electron mass. The Ce- and Yb- based intermetallic compounds, for instance, contain a band of strongly correlated $4f$-electrons whose $U/D$ is so large that they behave as local moments, along with a separate band of more weakly-interacting $s$, $p$, $d$-based conduction electrons. Another example are the iron based superconductors, containing electrons in partially occupied $3d$ orbitals, which show all the hallmarks of bad metals on the verge of Mott localization. Below, we shall discuss these two classes of materials in more detail.

**2.2. Quantum criticality in heavy fermion metals**

Quantum criticality often occurs in heavy fermion metals, intermetallic compounds containing rare-earth elements (such as Ce and Yb) with partially-filled $4f$- or $5f$-orbitals. The defining characteristics of the heavy fermion systems is the large effective mass of the electronic carriers. Its origin lies in the antiferromagnetic exchange interaction between the local $f$-electron moments and the conduction electron spins. This antiferromagnetic interaction leads the screening of local moments by the antiparallel spins of the conduction electrons. This is an example of a complex collective quantum behavior, known as the Kondo effect. As a result, the electronic excitations (Kondo resonances) grow out of the localized f -electrons; while itinerant conduction electrons end up having a small Fermi velocity $k_F/m^*$ or, equivalently, a large effective mass $m^*$. Correspondingly, the resulting electronic energy scales are small (of the order of $0.1 - 1$ meV), making it relatively easy to tune the electronic ground state by external parameters such as pressure or magnetic field.

**Figure 3**a illustrates the temperature-magnetic field phase diagram observed in the compound YbRh$_2$Si$_2$. At zero field, the compound develops an antiferromagnetic order at the Néel temperature, $T_N = 70$ mK. A quantum critical point occurs at the critical magnetic field, $B_c = 0.7$ T. Figure 3b demonstrates that while the quantum phase transition resides at T =0, quantum criticality affects a large temperature range, with the electrical resistivity being linear in T over three decades of temperature.

Microscopically, the heavy-fermion metals are described by a Kondo lattice model, comprising a lattice of localized magnetic moments and a band of conduction electrons. The conduction electron-mediated RKKY exchange interaction between the local moments drives them into a magnetically ordered state. Conversely, the Kondo-exchange interaction between the local moments and conduction electrons introduces spin flips, which tend to suppress the magnetic order and promote a Kondo-screened paramagnetic phase in which the local moments are screened by antiparallel spins of the conduction electrons. The two effects compete with each other, leading to a quantum critical point that separates the two phases, as mentioned in the introduction. The existence of the two competing interactions was already pointed out when the heavy-fermion materials were first discussed,[20, 21] but the discovery and active studies of heavy-fermion quantum criticality have only taken place over the past decade or so.[7-9]

Theoretically, how should one describe the collective excitations at a QCP? One way is to extend the theory of classical phase transitions, formulated by Landau,[22] which is based on the principle of spontaneous symmetry breaking. For magnetic systems, the spins are free to rotate in the disordered paramagnetic phase. In the antiferromagnetically-ordered phase, this continuous spin-rotational symmetry is spontaneously broken, as the spins must choose a preferred axis for their orientation. The Landau formulation characterizes the symmetry distinction in terms of an order parameter, which, for our case of an antiferromagnetic order, corresponds to the staggered magnetization. The order parameter gradually goes to zero as the critical point is approached, and the most important critical excitations are taken to be the spatial fluctuations of the order parameter. Generalizing the Landau paradigm to QCPs gives rise to essentially the same description.[23] The order-parameter fluctuations remain to be the primary critical collective excitations. The only manifestation of quantum mechanics is that the flucutations are not only spatial, but also temporal. For metallic antiferromagnetic systems such as heavy-fermion metals, what ensues is a SDW quantum critical point.[23-25]

Over the course of studying heavy-fermion QCPs, it has been appreciated that the Landau paradigm can break down for quantum phase transition. An alternative theory that emerged is that of a local quantum criticality.[26, 27] As advanced by one of the authors (Q.S.) and his collaborators, the local quantum criticality involves not only fluctuations of the magnetic order but also a critical destruction of the Kondo effect (**Fig. 4**a). Microscopically, the Kondo-exchange coupling not only destabilizes the magnetic order, but also introduces quantum entanglement between the local moments and conduction electrons. A smooth onset of Kondo entanglement yields its own quantum critical excitations. Because Kondo entangelement breaks no symmetry and is genuinely quantum mechanical, its involvement in differentiating the phases and in the quantum criticality makes the concept of local quantum criticality go beyond the Landau paradigm.

To expand on this, we note that the Kondo singlet represents a quantum-mechanical order (Fig. 4b); a singlet state is quantum entangled (as opposed to a product state). This description does not invoke a spontaneous breaking of any symmetry. When the antiferromagnetically ordered state breaks the Kondo singlet (Fig. 4c), the onset of the magnetic order is accompanied by a breakdown of the Kondo effect. The quantum criticality incorporates not only the slow fluctuations of the antiferromagnetic order parameter, but also the emergent degrees of freedom associated with the breakup of the Kondo singlet. This transition is referred to as locally critical,[26-28] because the antiferromagnetic phase transition is accompanied by localization of the $f$-electrons. This critical Kondo effect yields a new energy scale, $E^*_{loc}$, which continuously goes to zero at the QCP. The ensuing destruction of the Kondo resonances converts a large Fermi surface to a small one (Fig. 4a). Since the volume of the Fermi surface is directly related to the total number of the charge carriers by the Luttinger theorem, this implies that the effective number of charge carriers changes drastically across the quantum critical transition - a very non-trivial consequence!

There has been compelling experimental evidence for the Kondo-breakdown local quantum criticality, as summarized in a comprehensive review article.[7] This type of QCP provides an understanding of the unusual dynamical scaling properties observed in the inelastic neutron-scattering experiments, including the energy over temperature ($E/T$) scaling and a fractional exponent.[29, 30] Another salient property, predicted in the local quantum critical theory, is the new vanishing energy scale, $E^*_{loc}$ in Fig. 4a, which has subsequently been observed in YbRh$_2$Si.[31, 32] This appears as a temperature scale in the $T$-$B$ phase diagram, shown as the solid line in Fig. 3a. Yet another characteristic property predicted by the theory is a critically reconstructing Fermi surface across the magnetic QCP (Fig. 4a). Evidence for this has come from the Hall-effect and thermodynamic measurements in YbRh$_2$Si$_2$, as well as the de Haas-van Alphen (dHvA) measurements in CeRhIn$_5$.[33, 34]

Finally, to reflect back on the concept of emergent phases at the quantum critical point, there is considerable evidence that the Kondo breakdown local quantum criticality promotes superconductivity. The latter is found in proximity to the local QCP, with the superconducting $T_c$ that is very high compared to the Fermi energy of heavy fermions.[33-35] This is also corroborated by the involvement of the Kondo destruction in superconductivity, as inferred from inelastic neutron measurements of a nominally SDW QCP.[36] This issue is currently the subject of active theoretical studies.

## 2.3. Superconductivity and correlation effects in the iron pnictides and chalcogenides

High temperature superconductivity for a long time was the exclusive realm of the copper oxides.[37] This changed in early 2008, when superconductivity of $T_c \approx 26$ K was discovered in the doped LaFeAsO.[38] $T_c$ rapidly rose to about 55 K in the related 1111 iron pnictides.[39] High $T_c$ superconductivity has also been found in many other iron based compounds, notably the 122 iron pnictides such as the doped $BaFe_2As_2$,[40] 11 iron selenides such as FeSe,[41] and alkaline iron selenides such as $K_{1-y}Fe_{2-x}Se_2$.[42]

In these materials, superconductivity occurs when the antiferromagnetic order of a parent compound is suppressed by chemical doping. There is therefore a strong belief that the mechanism for superconductivity is connected to magnetism. Correspondingly, understanding the nature of magnetism in the parent iron pnictides and chalcogenides is central to the physics of these materials.

In the iron pnictides and chalcogenides, the electronic states near the Fermi energy are predominantly associated with the $3d$ orbitals of the Fe atoms. In the parent compounds, six electrons partially occupy the five $3d$ orbitals of each Fe atom. In order to understand their magnetism, we need to know the strength of electron correlations in these materials. As we mentioned in the context of Fig. 1, we can summarize the materials parameter space along the $U/D$ axis in terms of three regimes, which are specified again in **Fig. 5**a. When $U/D$ is small, the system corresponds to simple metals. Because of the Pauli exclusion principle, only electrons near the Fermi energy will be significantly influenced by the electron correlations. Both magnetic order and magnetic excitations are studied by perturbing around the non-interacting limit. When $U/D$ is very large, Mott localization sets in. In this Mott-insulating regime, electronic excitations are entirely "incoherent" (inherently many-body, rather than single-particle-like), forming the Hubbard bands. The spin degrees of freedom are contained in the localized moments, which are coupled to each other through the superexchange interactions. The intermediate values of $U/D$ give rise to the incipient Mott regime. Here, $U/D$ is below the threshold value for Mott transition, and the system is metallic. However, incoherent electronic excitations represent a significant portion of the single-electron spectral function; they represent the precursors to the Mott localization. These incoherent electronic excitations give rise to quasi-localized moments as part of the spin degrees of freedom.

The different regimes of electron correlations can also be specified in terms of the form of the single-electron spectrum. Fig. 5b shows the density of states of the single-electron excitations. The central part surrounding the Fermi energy ($E_F$) describes the low-energy coherent part. We use $w$ to label the fraction of the spectral weight in this coherent part. The remainder of the spectral weight lies in the incoherent part away from the Fermi energy, with a fraction $1-w$. Simple metals have a large $w$, while Mott insulators have $w = 0$. Bad metals that are incipiently Mott have a small $w$, with the majority of the electronic spectral weight lying in the incoherent part.

From early on, Si and Abrahams advanced the notion that the iron pnictides are bad metals, and that the dominant part of the spin spectral weight lies in the quasi-local moments with nearest-neighbor and next-nearest-neighbor ($J_1$-$J_2$) exchange interactions in the Fe square lattice.[44] Similar ideas have also been advanced by a number of other groups.[45-49] There is considerable evidence for this:

- The iron pnictide room-temperature electrical resistivity is very large, even when the residual resistivity is small, a hallmark of good material quality. The inferred meanfree path of quasiparticles at the room temperature would be on the order of the Fermi wavelength, as is typical for bad metals near Mott localization.

- The Drude spectral weight in optical conductivity[50] is significantly suppressed from its non-interacting counterpart. For $BaFe_2As_2$, for instance, the Drude weight is about 0.3 times of the non-interacting value,[50, 51] which implies $w \approx 0.3$. This provides a direct measure of the parent arsenide's proximity to the Mott transition. The smallness of the Drude weight is accompanied by the temperature-induced spectral weight transfer.[51-53]

- Zone boundary spin waves have been seen by inelastic neutron scattering (INS) measurements in several 122 iron pnictides compounds.[54] The spin spectral weight is large, and the spin damping is relatively small. Both features suggest the existence of quasi-localized moments, as expected in bad metals on the verge of a Mott transition.

We can tune the system further towards the Mott insulating side (decreasing $w$) by enhancing $U/D$. One way is to expand the Fe square lattice, which reduces the kinetic energy without causing much change to the on-site Coulomb repulsion. This is achieved in the iron oxychalcogenides, whose expanded Fe square lattice causes about 25% reduction in $D$ (from about 4.2 eV to 3.2 eV), thereby pushing the system to be a Mott insulator.[57] We can also reduce the kinetic energy by creating ordered vacancies on the Fe sites.[43] This is realized in alkaline iron selenides such as $K_{1-y}Fe_{2-x}Se_2$. The parent compound has an Fe layer with a $\sqrt{5} \times \sqrt{5}$ ordered vacancy pattern, with the usual 2+ Fe valence. It is a Mott insulator with a large ordered moment.[58, 59] **Fig. 6** demonstrates how the kinetic energy reduction associated with the ordered vacancies causes the enhanced tendency for Mott localization, using a simplified model in a simpler $2 \times 2$ vacancy ordering pattern as an illustration.

For the iron arsenides, the quasi-localized moments turn out to interact with each other through a nearest-neighbor exchange interaction, $J_1$, and a next-nearest-neighbor exchange interaction $J_2$, with $J_2$ that is comparable in magnitude to $J_1$.[44, 60] This provides a way to understand the ($\pi$, 0) collinear antiferromagnetic order that is observed in the parent iron arsenides.[61] This magnetic order is accompanied by an Ising order in the $J_1$-$J_2$ model,[62] which most naturally accounts for the tetragonal-to-orthorhombic distortion that is also seen experimentally.[61] Both orders are weakened by the coupling of the local moments to the coherent conduction electrons. When the coherent weight w is increased, this effect becomes more pronounced. Our detailed theoretical analysis, reported in Ref. [55], shows that increasing w will eventually suppress both the antiferromagnetic order and structural distortion, leading to a magnetic QCP, as illustrated in **Fig. 7**a. In Ref. [55], we also proposed that gradual P-doping for As represents a means to access this QCP. Subsequent experiments, both in Ce 1111 system (Fig. 7b)[56] and in Ba 122 system (Fig. 2b)[3, 63], have provided extensive evidence in support of this prediction.

The existence of parent compounds of both bad-metal and Mott-insulating ground states motivates the study of superconductivity in the doped systems using the Mott transition as the reference point. This leads to a 5-band $t$-$J_1$-$J_2$ model for studying superconductivity.[43, 64] $J_2$ promotes superconductivity in the extended $A_{1g}$ s-wave [$s(x^2y^2)$] channel, while $J_1$ favors that in the $B_{1g}$ d-wave channel. The magnetic frustration, represented by comparable magnitude of $J_2$ and $J_1$, therefore give rise to a quasi-degeneracy in the two channels.[64] This conclusion, also reached from weak-coupling calculations,[65] is of considerable phenomenological interest. For instance, the superconducting gap has been found experimentally to be either nodeless or nodal depending on the material class or even the doping level in a same family of iron based compounds. A particularly intriguing result of our strong-coupling studies is that superconductivity with strong pairing amplitudes not only occurs when the Fermi surface comprises both hole and electron pockets, but also arises when there are only electron Fermi pockets. This provides a natural understanding of comparably high $T_c$ observed in the alkaline iron selenides, which contain only electron Fermi pockets, as that seen in the iron pnictides, most of which contain both electron and hole Fermi pockets.

Here we have examined two classes of correlated materials, the heavy fermion systems and the iron pnictide high temperature superconductors. In the former, electronic interactions lead to the formation of local magnetic moments and quantum entanglement between those moments and conduction electrons. The competition between different orders, while purely quantum mechanical and tied deeply to zero-temperature quantum phase transitions, has ramifications over three decades in temperature through quantum criticality. Moreover, these ramifications can be profound, including a dramatic change in the size of the Fermi surface when crossing the transition. In the iron pnictides, the interaction-driven tendency toward localization coupled with the exchange interactions between the local magnetic moments have a profound impact on the occurrence of high temperature superconductivity. These systems both exemplify the rich physics and competing phases inherent in materials with strong correlations, and the way that the balance between orders can be altered by parameters such as pressure, doping, and magnetic field. The theoretical ideas discussed here showcase one approach to these complex systems, aiming to reveal general theoretical principles and unveil the essential physical ingredients required to understand their properties. This approach and others are discussed further in the next section.

**3. Electronic Structure in the Presence of Electronic Correlations**

A persistent challenge in materials physics, even more crucial in strongly correlated systems, is the need to find an appropriate theoretical description of real materials. Two families of approaches can be used to describe condensed matter systems theoretically: the *ab initio* (or first principles) electronic band theory, and the simplified model Hamiltonian approach (the power of which is shown in the previous section). These two approaches are often juxtaposed, as they start from quite different viewpoints. The *ab initio* approaches aim to accurately capture the underlying atomic and chemical structure of the material at hand, at the cost of having to make certain assumptions when treating the Coulomb repulsion between electrons. These approaches are thus well suited to study weakly correlated systems, such as "conventional" metals and insulators. The simplified model Hamiltonian approaches, on the other hand, strive to describe the essential electron-electron correlations as accurately as possible, at the cost of often oversimplifying the structural aspects of the material and dealing with a subset of all electrons (say, only *d*-electrons in transition metal compounds). These latter approaches hope to capture the essential physics qualitatively if not quantitatively.

The relative weaknesses of these two classes of approaches are evident. The fully *ab initio* methods based on density functional theory (DFT) typically fail to describe strongly correlated systems,[66] such as Mott insulators, whereas the model Hamiltonian methods rely on first finding an effective model for a given material, and the choice of the model and its parameters are often debatable. It is the authors' point of view that the successful theoretical understanding of strongly correlated materials hinges on intelligent combination of both types of approaches. One of the oldest known examples of such a combined methodology, still used to date, is the so-called DFT+U method [67] (also known as LDA+U, where LDA stands for "local density approximation"[68, 69]). One can also combine the *ab initio* electronic structure theory with more sophisticated numerical many-body approaches, such as the Dynamical Mean-Field Theory (DMFT),[70, 71] and since the initial proposal,[72] this promising methodology has seen many successful implementations (see Refs. [73, 74] and references therein).

In what follows, we shall provide several examples of how successful the combination of electronic structure theory and many-body approaches can be in describing correlated electron materials. In this section, we shall focus in particular on the heavy fermion *f*-electron material β-YbAlB$_4$ and the d-electron nickelate La$_4$Ni$_3$O$_8$ as two representative example. Later on, in Section 5, we shall also provide theoretical perspective on the metal-insulator transition in VO$_2$ and how it is affected by hydrogenation.

### 3.1. Heavy fermion metal β-YbAlB$_4$

In 2008, the Japanese team led by S. Nakatsuji announced the discovery[75] of a new ytterbium-based heavy-fermion material, β-YbAlB$_4$, whose atomic structure is shown in **Fig. 8**a. It consists of a planar network of boron atoms, with Yb and Al atoms sandwiched in between. This material has a very non-trivial low-temperature dependence of resistivity (the so-called non-Fermi liquid regime), $\rho = \rho_0 + \alpha T^{3/2}$, which becomes Fermi-liquid like ($\rho \sim T^2$) upon the application of magnetic field.[75] Most intriguingly, the effective mass of heavy quasi-particles appears to diverge as magnetic field $B \rightarrow 0$, indicating the breakdown of the Fermi liquid reminiscent of quantum criticality in other heavy fermion compounds, such as CeRhIn$_5$[76] and the aforementioned YbRh$_2$Si$_2$.[19, 77].

To gain a better understanding of this intriguing behavior, one of the authors (A.H.N.) has performed[78] *ab initio* band structure calculations of β-YbAlB$_4$, and found that just as expected from the experiment, the density of states at the Fermi level is dominated by narrow *f*-electron bands of ytterbium, as illustrated in Figs. 8b,c. The actual value of the calculated density of states at the Fermi level, $\gamma_{theor} \sim 6.7$ mJ/(mol(Yb)·K$^2$) falls short of the experimentally measured Sommerfeld coefficient in the applied field $\gamma_{exp}(B = 9T) \sim 60$ mJ/(mol·K$^2$), meaning that in this Fermi liquid regime, the correlations are very strong, with effective electron mass $m^*/m_{band} \sim 10$. As mentioned earlier, the effective mass gets even larger as the magnetic field $B \to 0$, and it is clear that the band theory is incapable of capturing this non-Fermi liquid behavior with diverging γ-coefficient.

Nevertheless, in spite of these issues, the ab initio density functional theory does prove useful in understanding certain important aspects of the material. For instance, the Luttinger theorem imposes a strict limit on the volume under the Fermi surface, and the *ab initio* predictions of the Fermi surface are therefore usually quite reliable, despite getting the effective mass of the carriers incorrectly. **Figure 9**b shows the calculated Fermi surface, with two large sheets and a smaller pocket in the Brillouin zone. To compare the Fermi surface with future de Haas-van Alphen experiments, the frequencies of the extremal orbits had been calculated, plotted in Fig. 9a. The low-lying frequencies characterized by relatively low effective electron mass (solid lines in Fig. 9a) turn out to correspond closely to the frequencies measured later in the de Haas-van Alphen experiment,[79] lending further credibility to the band theory calculations.

We note that the Fermi surface shown in Figure 9b includes the hybridization between the Yb *f*-electrons and the conduction electrons, so that in the language of Section , it corresponds to the so-called "large" Fermi surface on the Fermi-liquid side of the quantum critical transition. To experimentally access the "small Fermi surface" regime, one needs to tune the system to the ordered phase on the other side of the transition, which in the case of $\beta$-YbAlB$_4$ appears to be particularly challenging because no such ordered phase has been reported so far (although applying pressure may reveal a magnetically ordered phase). One also needs to bear in mind that the De Haas-van Alphen measurements require application of a sizable magnetic field, and this is problematic in the case of $\beta$-YbAlB$_4$, since an infinitesimally small magnetic field readily transforms the system into the Fermi liquid regime.[80]

The *ab initio* calculations, combined with model studies, have shed light on the origin of the heavy Fermi liquid state in β-YbAlB$_4$. One of the authors (A.H.N.), together with P. Coleman at Rutgers University, have constructed the minimal Kondo lattice model of this compound,[78] with the key ingredients coming from the *ab initio* calculations and the symmetry arguments. For instance, the band theory calculations showed that the overlap between electrons of Al and Yb is negligible, and that the largest contribution to hybridization comes from the hopping of B atoms on and off the Yb site[78]. Moreover, the local seven-fold symmetry of B atoms around the Yb site allowed us to draw important conclusions as to the local structure of the crystal field levels on the Yb site, deducing that the ground state doublet must be predominantly composed of $|J = 7/2, J_z = 5/2\rangle$ angular-momentum state.[78]

Further progress in understanding the physics of β-YbAlB$_4$ has been made by combining phenomenological theory with the magnetization measurements,[80] resulting in the realization of the field/temperature scaling of the thermodynamic properties, so that the free energy in the applied field $F(B, T)$ can be written in the following form:

$$F(B,T) \sim [(k_B T)^2 + (g\mu_B B)^2]^{3/4} \sim T^{3/2} f\left(\frac{B}{T}\right), \tag{1}$$

with $f(x)$ a universal function of the ratio $x = B/T$. This finding has far-reaching consequences for our understanding of the low-energy physics of the Kondo lattice problem in general and of β-YbAlB$_4$ in particular. It means that the response of the material to magnetic field and temperature is intimately linked, providing a fine example of the so-called universal scaling suspected to occur in the quantum critical regime of the heavy fermion systems.[7]

### 3.2. "Hidden order" in the nickelate La$_4$Ni$_3$O$_8$

Nickel is to the left of copper in the Periodic Table, with the electronic configuration [Ar]$3d^8 4s^2$. It is therefore conceivable that Ni$^{1+}$ ion may have similar electronic properties to Cu$^{2+}$, with a single hole in the otherwise filled $d$-shell: $3d^9$. Indeed, the discoverers of the high-temperature copper-oxide superconductors, Bednorz and Müller, initially investigated the possibility of superconductivity in La-Ni-O system[82]. However, at the time only compounds with Ni$^{2+}$ and Ni$^{3+}$ ions were known, and it is difficult to obtain the Ni$^+$ oxidation state. Theoretically, the LDA+U calculations of various nickel oxides showed that in order to achieve an $S = 1/2$ configuration in planar Ni-O configuration, Ni must indeed be in the rare Ni$^+$ state,[83] whereas the more common Ni$^{3+}$ state will not yield the necessary result. La$_4$Ni$_3$O$_8$ is a compound synthesized in the 1990s[84], but only studied in detail recently.[81] The compound has a nominal Ni valence Ni$^{+1.33}$ and incorporates two inequivalent Ni sites in a mixture of nominal Ni$^+$ and Ni$^{2+}$ oxidation states. La$_4$Ni$_3$O$_8$ belongs to the so-called T'-type series La$_{n+1}$Ni$_n$O$_{2n+2}$ ($n$ = 2, 3, ∞), which can be obtained by reduction of the Ruddleseden-Popper-type series A$_{n+1}$B$_n$O$_{3n+1}$.[84, 85] The structure of La$_4$Ni$_3$O$_8$ consists of alternating La/O$_2$/La layers and NiO$_2$/La/NiO$_2$/La/NiO$_2$ structural blocks.[81, 85]

Recently, Poltavets *et al.* have observed[81] a puzzling feature of La$_4$Ni$_3$O$_8$ at the temperature 105 K, which manifests itself as a jump in magnetic field-induced magnetization, a change of slope in resistivity (**Fig. 10**a) and a pronounced lambda-anomaly in the specific heat (Fig. 10b). All these features point to the onset of a thermodynamic bulk phase transition, yet its origin remained elusive at first. The samples are only available in the polycrystalline form, and both neutron powder diffraction and extended x-ray absorption fine structure (EXAFS) failed to detect any structural transition at 105 K.[81]

In order to clarify the nature of this "hidden order" phase below 105 K, one of the authors (A.H.N.) and Poltavets performed detailed first-principles calculations using the full-potential linearized augmented plane-wave method implemented in the Wien2k code.[86] Possible ordering, be it magnetic or charge in its origin, is complicated by the fact that there are three $NiO_2$ layers in the elementary unit cell, with two inequivalent atoms Ni(1) and Ni(2), as shown in **Fig. 11**a. Only the middle Ni(1)$O_2$ layer is truly planar with a square oxygen coordination, whereas the other two layers contain out-of-plane oxygen atoms. Because of the inequivalence between the two types of Ni atoms, it is conceivable that a type of charge ordering may take place. However, the calculations show that no such ordering occurs and that both Ni(1) and Ni(2) atoms are found in a mixed-valent $Ni^{1+/2+}$ state. This is corroborated by the results of neutron powder diffraction, which eliminate the possibility of significant charge ordering or of $Ni^{2+}$ high-spin to low-spin transition.

Significant correlations have been found between Ni $d$-electrons, with the screened Hubbard on-site repulsion calculated using the constrained-DFT method[87] to be U ≈ 4.3 eV. Using the DFT+U method, designed to capture at least to some extent the effect of strong electron correlations,[67] we found that the lowest energy ground state corresponds to spin-density wave (SDW) ordering of Ni $d$-electron spins with the ordering wave-vector **Q** = (1/3,1/3,0), which corresponds to a nine-fold in-plane increase of the magnetic unit cell, shown schematically in Fig. 11b.

The origin of this ordering can be tracked back to the Fermi surface geometry in the disordered, paramagnetic state. There are three quasi-two-dimensional Fermi surface sheets associated with Ni d-electrons. The first two are almost degenerate with each other and have the geometry shown in Fig. 11c, resembling the Fermi surface of the high-temperature cuprate superconductors. It is the third sheet, depicted in Fig. 11d, that is essential for the predicted SDW ordering because of the high degree of nesting between parallel parts of the Fermi surface. Quantitative analysis shows that the nesting wave-vector is very close to $\mathbf{Q} = (1/3,1/3,0)$, thus explaining the origin of the SDW ordering at this wave-vector. In the ordered phase, the Fermi surface reconstructs by opening up a gap along the third Fermi surface sheet, however according to the calculations, the density of states associated with the other two sheets in Fig. 11c remains non-zero. Thus theoretically, $La_4Ni_3O_8$ below its SDW ordering temperature is expected to remain a metal with a partially open gap at the chemical potential (known as the pseudogap). The fact that experimentally, the resistivity shows semiconducting behaviour, can be explained by the extreme Fermi surface anisotropy, so that the conductivity is metallic in the infinite $NiO_2$ planes and insulating in the direction perpendicular to those planes. This latter is speculated to be dominant in the poorly compacted polycrystalline samples,[81] thus explaining the observed semiconducting behavior (see Fig. 10a).

**3.3. Toward New Materials by Design**

As the above examples show, the first-principles calculations within DFT and DFT+U approaches have proven to be irreplaceable in helping determine the nature of the ordered phase below 105 K in $La_4Ni_3O_8$ and the electronic structure aspects of β-YbAlB$_4$. Even though the DFT and DFT+U methods and not always applicable to strongly interacting electron systems, we hope to have shown that their intelligent use and careful comparison with experiment prove to be extremely useful in understanding the nature of the low temperature quantum phenomena, such as heavy Kondo liquid in the case of β-YbAlB$_4$ and spin-density wave ordering in $La_4Ni_3O_8$. Further application of this philosophy to nanostructured materials will be discussed in Section 5.

As noted earlier, more sophisticated methods have been developed in recent years, combining *ab initio* electronic structure with for example, dynamical mean-field theory[71] within the LDA+DFMT framework.[72-74] These developments, combined with the availability of very large multi-processor computational resources, make the prospect of accurate prediction of electronic properties very exciting in the strongly correlated electron materials. At the same time, significant recent progress in materials synthesis and characterization makes it possible to envisage an integrated approach to new materials discovery, incorporating the materials growth, characterization, and ab initio computational modeling into a feedback loop.

New material discovery is a complex and lengthy process and it is highly desirable to have some guiding principles as a starting point for the materials synthesis. One such powerful principle is that of common structural "blocks" and more specifically, their particular structural arrangement into a periodic lattice. Below, we shall discuss an example of applying such a unifying structural principle to a very promising family of correlated electron materials, the transition metal dichalcogenides.

## 4. Transition Metal Dichalcogenides

Within the picture of strongly correlated materials, transition metal dichalcogenides $TX_2$ provide rich opportunities for exploring the emerging competition between various ground states. These materials constitute a class of layered compounds as versatile as they are complex. More than sixty such $TX_2$ systems have been known for decades, and yet their varied and often unexpected properties are yet to be deciphered.[88] Chemically, these materials consist of layers of early transition metal T and chalcogen X = S, Se or Te ions, arranged either as $TX_6$ edge sharing prisms or octahedra. The stacking of these layers along the orthogonal direction can also vary, and several polytypes usually exist for a given T and X combination. A wide spectrum of physical properties is represented in this rich class of materials, ranging from insulators ($HfS_2$), semiconductors ($MoS_2$; $TX_2$, T = Pt, Pd and X = S, Se), semi-metals ($WTe_2$, $TcS_2$) and metals ($NbS_2$, $VSe_2$).

Inherent to their nearly two-dimensional (2D) structure, the majority of the transition metal dichalcogenides display one or even multiple charge density wave (CDW) transitions.

These are periodic modulations of the density of electrons, brought about by the demand to minimize the global electronic energy when Coulomb repulsion is actually smaller than the energy gain from the gap opening as the new periodicity is introduced. Like the CDW, superconductivity (SC) is also a cooperative electronic state that occurs because of a Fermi surface instability and electron-phonon coupling. It is perhaps not surprising then, that CDW and SC often compete at low temperatures in a number of $TX_2$ compounds.[89-94] As a fundamental tenant of condensed matter physics, this competition has been intensely explored, and this has been facilitated mainly by the use of materials design as a tuning mechanism: electron doping (in $TaS_2$[95]), pressure (in $NbSe_2$,[96] $TaS_3$, $(TaSe_4)_2I$ and $NbSe_3$[91]) or both (in $Lu_5Ir_4Si_{10}$[94]) suppress the CDW transition to low temperatures, where the critical temperature of an existing superconducting state is consequently enhanced.

A new level of complexity, both chemically and physically, is achieved in these $TX_2$ systems, when various complexes M are intercalated[97] at the Van der Waals sites in-between the layers (**Fig. 12**), to form $M_xTX_2$. The structural and chemical changes associated with such intercalations drive the changes in physical properties, depending on the nature of the intercalant M:

(i) non-magnetic charge donor metals, such as Li, Na, K, or Ca, or Cu, Ag, Au, can be introduce at the M sites, and M often donates the electron to the conduction band. As a result, inter-layer interactions become considerable, such that the new structures are more three-dimensional. The corresponding changes in electron population provide grounds for the investigation of Fermi surface-driven transitions, and a way for tuning the electronic properties in a controllable way. $TiSe_2$ has proven to be an exceptional case, since the intercalation of either Cu[98, 99] or Pd[100] has unveiled surprising structural changes with $x$ (a counterintuitive increase of the layer spacing), as well as a novel SC state at finite x, where charge doping alone does not explain the dependence of the superconducting temperature on $x$.

(ii) the opposite structural effect is achieved as a result of intercalating organic molecules (typically large) at the M site. Up to an order of magnitude increase in inter-layer spacing can be achieved when M is a long chain amine, or pyridine,[101] and the properties of nearly single layers can be studied individually.

(iii) the most notable effect of the intercalation of 3d transition metals M = V, Cr, Mn, Fe, Co or Ni is that of a magnetically ordered ground state in the resulting $M_xTX_2$ compounds. While the magnetism greatly depends on the choice of both M and T transition metals,[102] the amount of intercalant also dictates the structural and electronic properties.[103] Remarkably, for T = Nb or Ta, and M = V, Cr, Mn, Fe or Co, hyperstructures are stabilized for $x = 1/4$ or 1/3 amount of intercalant. Antiferromagnetic order sets in at low temperatures in $Fe_{1/3}NbSe_2$, $Fe_{1/4}NbSe_2$, $Co_{1/3}TaS_2$ or $Ni_{1/3}TaS_2$, while the majority of other complexes order ferromagnetically. Enhanced magnetic moments are observed in $Mn_xNbS_2$, and this has been attributed to the hybridization between the local moments and the conduction electrons. A most unpredictable magnetic behavior was observed upon Fe intercalation in $TaS_2$, where, for $x = 1/4$, sharp switching of the magnetic domain orientation occurs at high magnetic fields (around 4 T) and relatively high temperatures (up to 100 K).[100]

Our group at Rice continues to look at some most outstanding intercalated layered dichalcogenides, with emphasis on the drastic changes in structural, electronic or magnetic properties as a result of the intercalation. This overview will likely reinforce the idea that no unifying theory exists, which could account for the complex phase transitions accompanying the underlying *d*-electron physics in the parent (non-intercalated) compounds. In turn, the search for such a theory is even more justified by the need to make the synthesis of target compounds more predictable and controllable.

## 4.1. The competition between Charge Density Wave and Superconductivity: A comparison between $Cu_xTiSe_2$ and $Pd_xTiSe_2$

Even without any intercalation, TiSe$_2$ is an intriguing material, one of the earliest discovered layered dichalcogenides. Despite its apparent chemical simplicity - it only forms in one polytype known as the 1T form, which means that the unit cell is trigonal (T) and consists of a single (1) TiSe$_2$ layer - its electronic properties have been controversial and not well-understood for decades. This compound undergoes a transition to a commensurate CDW (CCDW) state below 220 K,[104] when it is more common for a incommensurate CDW (ICDW) to occur at high temperatures, with the CCDW is stabilized only upon further lowering the temperature. CDW transitions are favored by Fermi surface nesting - the presence of large parallel Fermi surface sheets; however this is not the case in TiSe$_2$, and the CDW mechanism has been attributed to different scenarios, including an antiferroelectric mechanism,[105] exciton formation,[106, 107] or a indirect Jahn-Teller effect.[108] Even the normal state electronic properties had been subject to a long lasting debate, with many studies - spanning nearly three decades - arguing for either a semiconductor[109-114] or semimetal[115-119] state above the CDW transition in TiSe$_2$. The difficulty in settling this argument was due to the small indirect gap present in this compound, which could not be accurately resolved as positive or negative.

Renewed interest in this compound was brought about by the recent discovery of SC induced by Cu intercalation,[98] and the current, more advanced experimental tools (*e.g.*, optical spectroscopy) have now concluded that the compound is metallic in both the normal and CDW phases.[120] As an electron donor, Cu would be expected to bring the TiSe$_2$ layers closer together, as is the case with most non-magnetic intercalants.[121-124] Surprisingly, a monotonic expansion of the unit cell is observed with increasing amounts $x$ in Cu$_x$TiSe$_2$,[98] until the solubility limit is reached around $x = 0.11$. With increasing $x$, the CDW transition is suppressed, and superconductivity occurs for $x \geq 0.4$, even before the CDW transition completely disappears. It should be pointed out that the CDW-to-SC competition observed in Cu$_x$TiSe$_2$ is unique among layered dichalcogenides, since no SC exists in the parent compound. This made $Cu_xTiSe_2$ the first example of such a transition induced by chemical tuning. Since this discovery, we found that superconductivity can also be induced in $TiSe_2$ by Pd intercalation,[100] while notable similarities and difference exist between the two compounds:

With both Cu and Pd intercalation, the inter-layer spacing c increases monotonically when $x$ increases. Perhaps correlated with this lattice expansion is the occurrence of superconductivity when $c \approx 6.03 \pm 0.01$ Å; conversely, no superconductivity is observed when the intercalant shrinks the $c$ lattice constant.[121-125] However, this apparent correlation between the structural properties and superconductivity hardly lends itself to generalization, as there are also substantive differences in the CDW-to-SC transition in the two compounds. With increasing $x$ in $Cu_xTiSe_2$, a monotonous increase is observed for the electronic specific heat coefficient γ and the room temperature magnetic susceptibility χ(300K) (**Fig. 13**a).

The same parameters γ and χ(300K) (Fig. 13b) actually are at a maximum around the SC state in the case of $Pd_xTiSe_2$, and decline for larger $x$. While Cu and Pd are expected to contribute one and zero electrons, respectively, when intercalated in $TiSe_2$, it is not straightforward to understand the nearly double maximum superconducting temperature $T_c$ = 4.15 K in $Cu_xTiSe_2$ at $x_c \approx 0.08$ (Fig. 15a), compared to the corresponding temperature $T_c \approx 2$ K in $Pd_xTiSe_2$ at $x_c \approx 0.11$ (Fig. 15b). A striking effect of Pd intercalation on the transport properties of $TiSe_2$ is that a single tuning parameter (chemical doping in this case) drives the system from a semimetal ($x = 0$), to a insulating ($0.01 < x \leq 0.06$), to a metallic ($0.08 < x < 0.10$), to a superconducting ($0.11 \leq x \leq 0.12$), and finally to a metallic state ($x > 0.12$) (**Fig. 14**). Adding to the complexity of the electronic state in $Pd_xTiSe_2$ is the presence of a second CDW transition close to the emerging insulating state $0.01 \leq x \leq 0.03$; although multiple CDW transitions have been previously observed in other chalcogenides,[126-128] oxides[129] or even intermetallics,[130] $Pd_xTiSe_2$ is the only system so far where a second electronic transition is induced within an already CCDW state.

Competing interactions can result in quantum criticality, and this is indeed apparent in $Cu_xTiSe_2$. With a $T$-$x$ phase diagram reminiscent of that of the high temperature cuprate superconductors (**Fig. 15**a), the CDW phase boundary in $Cu_xTiSe_2$ intersects the superconducting dome around $x = 0.06$.[98] Raman scattering studies[99] suggest that the CDW phase boundary extends into the superconducting state, and results in a quantum critical point close to $x = 0.07$. This observation implies a further correlation between the quantum critical behavior and the lattice expansion in $Cu_xTiSe_2$: by contrast with Cu intercalation, where the lattice expands with $x$, pressure effects on $TiSe_2$[131] lead to lattice contraction, with no sign of the softening of the CDW towards quantum criticality. The natural question then arises, whether the lattice expansion in $Pd_xTiSe_2$ would also result in a quantum critical point in this compound, and the $T$-$x$ phase diagram (Fig. 15b) suggests that this may be the case around a critical composition $x = 0.10$.

## 4.2. Sharp magnetic domain switching in $Fe_{1/4}TaS_2$

Fe-intercalated $TaS_2$ captures several of the most important effects of chemically tuning the layered dichalcogenides with magnetic ions, while also covering a very rich range of electronic and magnetic properties, not easily accounted for by existing theories. With this overview of the properties of $Fe_xTaS_2$, we reemphasize the outstanding versatility of the chemical doping in the $TX_2$-type materials, but also the lack of unifying theoretical principles to guide the design of novel systems with targeted properties.

The magnetic and transport properties of $TX_2$ compounds, when intercalated with 3d transition metals M = Cr, Mn, Fe, Co or Ni, have revealed[102, 103] unusual magnetic ground states, where both localized and itinerant magnetic moments coexist. Such behavior can be attributed to the nearly two-dimensional crystal structure, in which the local moments of the M ions are separated from the conduction electrons within the $TX_2$ layers, rendering their properties akin to those of the rare earth metals. In particular, the Fe intercalated $TX_2$ system illustrates the complexity of the ground state properties in $Fe_xTaS_2$ systems:

for T = Ta and $x$ = 1/4 or 1/3, a ferromagnetic (FM) ground state is realized in a 2 × 2 or √3 × √3 superstructure respectively. The most surprising property resulting from the Fe intercalation is the non-monotonic change of the Curie temperature $T_C$ for 0.2 ≤ x ≤ 0.4, such that $T_C$ = 160 K for $x$ = 1/4, and only 35 K for $x$ = 1/3.[132] Furthermore, the magnetic ground state changes from FM for $x$ < 0.4 to antiferromagnetic (AFM) for 0.4 < $x$, or from FM for $x$ = 1/3 and T = Ta to AFM for $x$ = 1/3 and T = Nb.[88] It is readily apparent that the intercalated layered dichalcogenides offer a rich class of materials for the search of novel magnetic properties.

$Fe_{1/4}TaS_2$ stands out among these systems, due to its remarkable magnetic ground state, which, in turn, may lend itself to potential applications. $Fe_{1/4}TaS_2$ displays an unusually sharp switching of the magnetization $M$ as a function of field (**Fig. 16**), with a high switching field $H_s$ ≈ 4 T, and a large saturated magnetic moment $\mu_{sat}$ = 4 $\mu_B$/Fe.[133] The nearly square hysteresis loops are observed in this compound for temperatures up to 100 K, with the magnetic moment per Fe nearly field- and temperature-independent. Electron diffraction images parallel to the (001) direction (**Fig. 17**a) show the presence of sharp reflections corresponding to the basic trigonal structure and also the 2$a$ superstructure reflections; this indicates that the Fe moments are indeed ordered in the hexagonal $ab$ plane. However, the superreflections corresponding to the direction perpendicular to the planes (Fig. 17b) are streaked, suggesting that there is disorder in the stacking of the planes along the $c$ axis. When variable amount of Fe was intercalated in $Fe_xTaS_2$, no other composition besides $x$ = 1/4 showed the disordered stacking of the hexagonal planes or the sharp $M(H)$ isotherms,[134] suggesting a potential correlation between the structural order and the magnetic ground state.

Additional details of the magnetic anisotropy in $Fe_{1/4}TaS_2$ can be inferred from the electrical resistivity measurements, as well as from magneto-optical (MO) imaging.[135] The in-plane resistivity (**Fig. 18**a) shows weak temperature dependence at temperatures above $T_C$, with a decrease at the ferromagnetic ordering, consistent with less scattering as the magnetic moments align at lower temperatures. Given the leveling-off of the resistivity at low temperatures, it is surprising that the domain switching time τ decreases linearly with $T$ (inset, Fig. 18a), suggesting that the eddy currents cannot be the major factor in determining the switching time.

While the $H \parallel ab$ magnetoresistance is quadratic in $H$, the $H \perp ab$ magnetoresistance has a remarkable field dependence (Fig. 18b) which qualitatively reproduces the features observed in the corresponding $M(H)$: a sharp drop in $\rho(H)$ occurs at $H_s$, with the magnetoresistance changing nearly linearly with H away from $H_s$. Further evidence for the magnetic anisotropy in $Fe_{1/4}TaS_2$ is provided by the MO images in **Fig. 19**: field cooling of the sample resulted in a magnetic monodomain for the whole measured crystal (Fig. 19, top left), and increasing magnetic field resulted in unusual dendritic nucleation, with the dendrites parallel to the crystallographic a and b axes. This is suggestive not only of large c-axis anisotropy, but also significant anisotropy with respect to domain-wall motion within the hexagonal plane. An additional study of the radio-frequency magnetic susceptibility $\chi_{RF}$[135] provided strong evidence for local moment FM in this system. Subsequent ARPES experiments and band structure calculations[136] had shown a large unquenched orbital magnetic moment (~ 1 $\mu_B$ /Fe), in addition to the large local moment close to 4 $\mu_B$ /Fe, as seen from magnetization.

It should be readily apparent, even from these few distinct examples of intercalated dichalcogenides, that they represent a fascinating family of correlated materials. The ubiquitous charge density wave ground state in most parent compounds $TX_2$ is challenged by either other electronic states (superconductivity) or magnetism when intercalants are inserted in-between the dichalcogenide layers. While, for the most part, the electronic properties of these compounds end up closer to the metallic (small to moderate $U$) range of the correlation strength, the emerging ground states and the accompanying transitions are intriguing none the less. Quantum phase transitions occur in some cases within a dome of superconductivity, as induced by the amount of intercalant *x* in $M_xTiX_2$, when M is non-magnetic; or, strikingly, unusual magnetic behavior can result in some case, possibly due to a combination of structural in-plane spin order, and inter-plane short range order (or disorder) in the case of magnetic M, and for specific amounts of *x*. An overarching picture of such complex behavior may only be achieved with both the development of more such materials exhibiting similar effects, and with theoretical input which, in turn, can be then implemented at the materials design stage.

## 5. Nanostructures to Examine Strongly Correlated Materials

The last twenty years have seen a dramatic growth in the availability of nanofabrication capability, particularly electron-beam lithography, and nanoscale characterization methods, such as scanned probe microscopy and transmission electron microscopy. At the same time, there has been increasing recognition that structuring materials on the nanoscale can profoundly modify material properties and enable experiments that are otherwise not possible using macroscopic samples. Most of the nanoscale attention has focused on traditional semi conductors, metals, and carbon nanomaterials, largely for reasons of historical development, compatibility of materials with processing conditions, and technological motivation.

However, there is a growing appreciation that there is much to be gained by applying these nanoscale methods to strongly correlated materials.[137] As described above, in such systems there is often an extremely close competition between electronic (and structural) states with dramatically different properties, the outcome in equilibrium being driven by kinetic, Coulomb, magnetic, and structural energy scales. Transitions between competing phases can exhibit both intrinsic and extrinsic inhomogeneities over various length scales. Incorporating strongly correlated materials directly into nanostructure-based experiments can give access to these phases at the single domain spatial level. Nanostructured materials, with their comparatively large surface-to-volume ratios, may also be modified in ways not generally possible in bulk systems, such as through electrostatic gating in a field-effect geometry, or through surface-mediated intercalation of chemical dopants. Electrodes with nanoscale separations allow the application of large perturbing electric fields with comparatively moderate voltages. The ability to inject and remove current and sense voltage on the nanoscale sets the stage for non-equilibrium studies of strongly correlated systems, with the electronic and ionic degrees of freedom taking on steady state but non-thermal distributions.

At Rice we have so far focused our nanostructure-based studies primarily on two materials systems, magnetite ($Fe_3O_4$) and vanadium dioxide ($VO_2$). Both are strongly correlated transition metal oxides, in which the localized character of partially occupied *d* orbitals leads to electron-electron interactions taking on an enhanced importance. Both materials also have comparatively strong couplings between electronic degrees of freedom and optical phonons that grow soft. As a result, each system shows a particular kind of metal-insulator transition (MIT) coupled with a structural transition to an atomic arrangement of reduced symmetry as the temperature is reduced below a critical value. We discuss each of these materials and our experiments on them in turn.

### 5.1. $VO_2$ and gating

Vanadium dioxide exhibits a dramatic first-order phase transition at 67°C between a high temperature, metallic, rutile structure and a low temperature, insulating monoclinic structure with an energy gap of 0.6 eV as inferred from the activation energy of the resistivity. As $VO_2$ is cooled through the transition, the electronic conductivity drops by four orders of magnitude. This large change in electronic (and optical) properties and its proximity to room temperature suggest that this material may be of interest for a number of applications, and the ability to tune or control the transition is of much interest.

While the vanadium ions formally each have a single unpaired *d* electron, there is no magnetic ordering in this system, which remains a paramagnet over the whole temperature range down to cryogenic conditions. There has been a long-standing controversy concerning the mechanism of the MIT in $VO_2$. On the one hand, conduction in the metallic state is moderately correlated[138] (the *w* parameter[50] described in Fig. 5 is ~0.5) and involves the half-filled *d*-band, suggestive of Mott physics. On the other hand, the structural transition into the monoclinic state can be viewed as dimerization of vanadium chains, hinting at a Peierls character to the transition. A consensus is emerging that the MIT is a correlation-enhanced Mott-Peierls transition, with both the on-site repulsion of the unpaired *d* electrons and the strong coupling to lattice degrees of freedom playing critical roles. Because of the difference in structures of the two phases, strain has a profound effect on the transition. In the coexistence regime, $VO_2$ will spontaneously break up into metallic and insulating domains, with nanoscale measurements able to probe at the single domain level.[139]

Interest in $VO_2$ has undergone a resurgence in recent years, with the growing realization in the community that it may be grown readily as single-crystal nanowires.[140] In nanowire form, it has proven possible to examine the MIT in great detail, revealing much about the phase diagram in the presence of strain.[141-144] Doped nanowires, with tungsten substituted on the vanadium sites, have also been considered.[145, 146]

Nanostructures make possible experimental approaches that try to alter the balance between the competing metallic and insulating phases. One avenue that has been pursued lately with good success is the electrostatic gating of strongly correlated materials. In the field-effect geometry, a gate voltage is intended to tune capacitively the carrier density in a two-dimensional channel region between source and drain electrodes. In strongly correlated materials, the carrier density typically must be altered by a significant fraction of a charge carrier per unit cell to be in the same regime as the chemical doping techniques used to modify bulk materials. Over the last few years, ionic liquids have been employed as electrolytic gate media, achieving surface charge densities exceeding $10^{14}$ carriers per $cm^2$. As a result, semiconductors have been gated into metallicity,[147, 148] and insulating parent compounds of high temperature superconductors have been gated into the superconducting regime.[149-151]

Repeated attempts have been made to gate the MIT in $VO_2$ films, with the intent of developing a "Mott-FET". Gating in the insulating state with conventional gate dielectrics has shown little response in the conductance,[152, 153] with no sign of a gate-driven MIT. Yang *et al*.[154] employed an ionic liquid gating configuration on a thin film $VO_2$ device, finding minimal, slow, hysteretic gate response at positive gate biases.

As reported recently,[155] we have performed similar experiments using individual $VO_2$ nanowires as the channel material rather than sputtered films. The nanowire geometry, with its large surface to volume ratio, favors this approach. Consider a typical $VO_2$ nanowire 200 nm wide, 100 nm thick, and 10 μm in length. A typical carrier density in the insulating phase near room temperature is $5 \times 10^{18}$ cm$^{-3}$, or approximately $10^6$ carriers total. For comparison, such a nanowire sitting on a substrate has an exposed surface area of $4 \times 10^{-8}$ cm$^2$. A conservative gated carrier density in an ionic liquid gate structure is $10^{14}$ carriers/cm$^2$, implying that it should be readily possible to modulate the carrier population in the nanowire by an amount comparable to the total thermal population.

Our results are summarized in **Figure 20**. Our nanowire devices have the rutile *c* axis (the monoclinic *a* axis) oriented along the long axis of the nanowire. Source/drain contacts are made from Au with a thin (5 nm) Ti adhesion layer. Conductance is measured through standard lock-in methods at a frequency of 13 Hz. Thermal cycling in the absence of any gate bias shows the usual MIT, with a conductance change of four orders of magnitude between the insulating and metallic states. The ionic liquid in these devices is DEME-TMSI, known for its large window of electrochemical stability.

As the gate is cycled to positive voltages and back at room temperature (sweep rate 2 V/min) and below (well in the insulating state), we observe no detectable change in the source-drain conductance. This is surprising in light of the simple estimation above. At the very least, it suggests that treating VO$_2$ in the insulating state as if it is a conventional semiconductor is unwise. It would appear that there are two potential explanations. First, if the density of localized surface states is somehow exceptionally large, then perhaps the gated charge is insufficient to create a significant change in the number of mobile carriers. Given the success had in ionic liquid gating of other oxide surfaces and the single-crystal character of the nanowires, this seems unlikely. The other alternative is that the field from the ionic liquid is not fully screened by the carriers in the nanowire. If achieving the required carrier density for screening would necessitate the structural transition to the rutile state, perhaps the strain energy cost is prohibitive. Additional measurements (*e.g.*, in wider nanowires in a Hall configuration, or coupled with capacitive measurements to look at screening dynamics) should be helpful in determining between these alternatives.

In the course of our gating measurements, we found that the presence of trace amounts of water in the ionic liquid led to slow, hysteretic gate response. In particular, at elevated temperatures, biasing the gate to positive values (e.g., +1 V) while the device was in the metallic state resulted in a persistent change (even with the gate voltage reduced back to zero) in the properties of the subsequently measured insulating state. This persistence echoes the trends seen previously[154] in sputtered VO$_2$ films, and is consistent with an electrochemical doping process.

The dopant in question appears to be atomic hydrogen. Hydrogen diffuses readily within the rutile structure of TiO$_2$[156] and acts as an electron donor in that material.[157] Doping of VO$_2$ with hydrogen was first reported over twenty years ago,[158] though not explored in detail. More recently, electrochemical doping of sputtered VO$_2$ films has been demonstrated,[159] though in these situations strain makes it challenging to get a clear picture of the intrinsic properties of the doped material. There is a report that electrochemical doping can lead to a metallic state in the monoclinic structure.[160] Conversely, starting from water-containing paramontroseite, Wu *et al.*[161] have shown that thermal decomposition can produce hydrogen-doped $VO_2$ that appears to be in a stable rutile structure at and below room temperature.

We have investigated hydrogen doping of $VO_2$ at the level of single nanocrystals,[162] using catalytic spillover from lithographically defined metal electrodes as the source of atomic hydrogen. Upon exposure to a partial pressure of hydrogen under very modest conditions (e.g., $150°C$ for tens of minutes) and then cooled to room temperature, nanocrystals in contact with catalytic metal (Pd, Au, Cu, Ni) pads show obvious changes observable in an optical microscope, taking on a color very similar to that seen in undoped $VO_2$ material when heated into the metallic rutile state (**Fig. 21**). Nanocrystals on the same substrate not in contact with catalytic metal show no changes under these conditions. Detailed examination via Raman microscopy shows that the doped material has a Raman spectrum dominated by continuum emission, as in metallic, rutile $VO_2$. Structural characterization via transmission electron microscopy (TEM) and selected area electron diffraction (SAED) confirm that at high doping the nanowires remain crystalline upon doping and adopt a configuration consistent with a rutile structure stabilized at room temperature.

Electronic transport measurements show that moderate hydrogen doping increases the conductivity and reduces the activation energy describing the conductivity in the insulating state, while changing the transition temperature little. When doped maximally, the transition vanishes entirely, and metallic conductivity is stabilized down to cryogenic temperatures. Remarkably, this doping process is completely reversible. Baking the doped device in air at $250°C$ for 20 minutes returns the material to a state with electronic and optical properties that appear identical to that of undoped $VO_2$.

Having demonstrated that it is possible to manipulate the MIT and tune the properties of $VO_2$, it is natural to ask what mechanism is at work. To do this we turn to electronic structure calculations as a guide. Conventional density functional theory (DFT) is challenged by $VO_2$. Properly incorporating the on-site repulsion relevant for the d electrons is a challenge while also avoiding an erroneous prediction of DFT+U that the material should order antiferromagnetically at low temperatures (experimentally, this only happens under uniaxial strain). Nevertheless, the structural properties tend to be predicted by DFT quite reliably even for strongly correlated Mott insulators such as MnO and NiO.[163, 164] The first question that needs to be answered is how hydrogenation affects (if at all) the structure of $VO_2$, and in this respect, the DFT calculations are expected to be reliable.

In the experiment, hydrogen catalytically penetrates into the host material through the surface, and the actual hydrogen concentration likely has very complicated non-uniform dependence on the depth below the sample surface as well as the distance from the catalyst. To make the calculations feasible, we made a simplifying assumption that hydrogenation occurs uniformly in the bulk, which is likely to be correct for the finely crashed $VO_2$ powder is exposed to hydrogen. In the calculations, the hydrogen concentration can be varied by constructing a supercell consisting of multiple $VO_2$ units, for example 25 mol% would correspond to the chemical formula $HV_4O_8$, with one hydrogen atom per 4 formula unit cells of $VO_2$. We have studied a few hydrogen concentrations (between 12.5 mol% and 100 mol%), using a variety of initial hydrogen positions and letting the structure "relax" by minimizing the total energy of the system. Invariably, we found that the hydrogen atoms penetrate the structure along the open channels in the rutile structure along the c axis (perpendicular to the plane of **Fig. 23**) and form a strong bond with one of the oxygen atoms, as shown in Fig. 23 (the same mechanism, it turns out, was established earlier for hydrogen-intercalated $TiO_2$ [166]). Responding to hydrogenation, the rutile structure of metallic $VO_2$ distorts, as depicted in Fig. 23, and the lattice expands. The simulated diffraction pattern associated with this theoretical structure agrees well with the electron diffraction data.[162]

The most important question, of course, is why does hydrogenation favor the metallic (pseudo-)rutile structure over the insulating monoclinic one. Both the DFT and DFT+U methods predict a range of concentrations over which the pseudo-rutile structure is thermodynamically more stable than the monoclinic one at absolute zero temperature,[162] corroborating the experiments. From the chemistry point of view, this can be understood by considering that the H..O bond formation will lead to electron transfer from hydrogen onto oxygen atom, therefore reducing the electronegativity of the latter. As a result, the V..O bond will become less ionic, meaning that less electrons will be donated by vanadium atom, making V $d$-level more populated and hence resulting in a more metallic behavior. This schematic argument can be tested rigorously with first-principles calculations, and we found that indeed, the vanadium $d$-orbital occupancy rises monotonically with hydrogen concentration, up to as much as 0.16 electrons for 100 mol% (in $HVO_2$). Since the density of states at the Fermi level comes primarily from V $d$-electrons, this translates directly to doping the system away from half-filling and from Mott metal-insulator transition, which is what the experiments demonstrate.

Another factor that plays a role in the metal-insulator transition is the dimerization of neighboring V atoms along the rutile c-axis, which occurs below the transition into the monoclinic phase. Such behavior, typical of one-dimensional systems such as polyacetylene,[167] opens up a gap at the Fermi surface and is known as the Peierls mechanism.[168] In fact, the earlier LDA+DMFT calculations on pristine $VO_2$ indicate that the metal-insulator transition is not of pure Mott type, but is rather interaction-assisted Peierls transition.[169, 170] From our calculations, we find that the unit cell volume expands by as much as 13% on hydrogenation, meaning that the overlap between electron shells of neighboring vanadium atoms will be reduced as it depends exponentially on the interatomic distance. This in turn would weaken the energy gain for Peierls distortion, which is proportional to the hopping matrix element between the atoms.[168]

From the above discussion, we conclude that hydrogenation of $VO_2$ both dopes the $d$-electron orbitals away from half-filling and at the same time reduced the tendency towards Peierls transition, thus explaining qualitatively the experimentally observed suppression of the metal-insulator transition in VO$_2$ nanobeams upon hydrogenation. We arrived at these qualitative results from theoretical DFT and DFT+U calculations, despite known concerns about applicability of these methods to quantitative description of strong electron correlations. While more sophisticated methods such as GW, DFT+DMFT and particular hybrid functionals are known to yield better results for pure VO$_2$,[169, 171, 172] they are computationally prohibitively expensive to deal with the large multiformula-unit supercells necessary to treat hydrogenated VO$_2$. The above is an example of a fruitful symbiosis between experiment and theory applied to intrinsically very complex correlated system with competing electron ground states.

This work is an example of the utility of nanostructures and nano-based techniques in improving our understanding and ability to control strongly correlated materials. The diffusion of the hydrogen is best examined on micron scales, with the individual nanowires acting as perfect test subjects for studies under comparatively minimal strain. Achieving comparative homogeneity of sample properties is easier in nanocrystals than in a macroscale crystal or powder. As a result, we have shown, both experimentally and theoretically, that catalytic spillover as a means of hydrogen doping can be used to tip the equilibrium balance between competing correlated ground states.

### 5.2. Magnetite driven out of equilibrium

In contrast, our work on magnetite (Fe$_3$O$_4$) takes advantage of nanostructure techniques to perturb the material out of equilibrium, as a means of examining the stability of the correlated ground state. Magnetite at room temperature is a ferrimagnetically ordered material with an inverse spinel crystal structure with a point group symmetry Fd3m and a comparatively large unit cell (lattice parameter a = 0.8396 nm). The minority spin A-site iron atoms are tetrahedrally coordinated by oxygen atoms and have a formal charge of +3. The majority spin B-site iron atoms are octahedrally coordinated by oxygen, and have a formal charge that is distributed 50/50 between +2 and +3. In this high temperature state magnetite is moderately conductive, with a resistivity of a few milliOhm-cm, with conduction thought to take place via fluctuating valence of the B-sites. Since the material is ferrimagnetically ordered below $T_C \approx 860$ K, the mobile carriers are expected to be fully spin-polarized. Upon cooling, the resistivity increases with an activation energy comparable to 25 meV, showing that this state is not a conventional metal. Magnetite undergoes the so-called Verwey transition[173] at 122 K, a first order transition to an orthorhombic crystal structure of symmetry P2/c, with a resistivity 1-2 orders of magnitude larger and a more insulating temperature dependence.

The nature of the magnetite ground state and the physical mechanism of the Verwey transition have been controversial for decades. Originally conceived as a charge ordering transition, the actual situation is considerably more complex,[174-178] involving a combination of lattice distortion, charge disproportionation, and orbital ordering. Experiments further indicate an onset of ferroelectric polarization below 40 K.[179] As in vanadium dioxide, attempts to model the transition require the inclusion of both strong on-site repulsion of the d electrons and a strong coupling of carriers to the lattice.

Our work in magnetite springs from the discovery[180] that the application of a strong dc electric field can kick the system out of the insulating Verwey ground state and into a more conducting state (**Fig. 24**). Qualitatively similar non-equilibrium transitions have been reported previously in strongly correlated materials. For example, Asamitsu *et al.*[181] demonstrated that the nominally charge-ordered, insulating ground state of mm-sized crystals of $Pr_{1-x}Ca_xMnO_3$ can be disrupted by the application of a dc bias of several hundred volts, and that the insulating state is restored upon the removal of the bias, though hysteresis is observed as the voltage is swept up and back. Similar breakdowns of correlated states have been reported in other correlated insulators.[182-184]

By fabricating closely spaced electrodes on the surface of epitaxially grown magnetite thin films (50-100 nm thickness), it is possible to apply large in-plane electric fields at modest voltages. Using a self-aligned process,[185] electrodes of appreciable width (~10-20 μm) can be separated by less than 10 nm. The linear scaling of the threshold switching voltage with interelectrode spacing on a single substrate helps to confirm that the breakdown process is driven by electric field rather than carrier energy. The typical electric field required to destabilize the insulating state is approximately $10^7$ V/m at 80 K. This critical field decreases as $T$ is increased, but does not approach zero as $T \to T_V$. Small electrode separations make possible switching voltages on the order of a volt, considerably below the scale where carriers have sufficient energy to do chemical damage to the material as in "hard" breakdown of dielectric oxides.

The bias-driven breakdown was only observed in samples that showed a well-defined Verwey transition, and only at temperatures below $T_V$. It was of critical interest to determine the role of self-heating in our experiments. Continuous sweeping of the dc voltage led to significant hysteresis in the $I$-$V$ characteristics. However, if instead the conductance was investigated using pulsed voltage techniques to minimize self-heating, the hysteresis could be greatly reduced, leaving a well-defined switching voltage, $V_{sw}$ for a given device at a given temperature. Modeling these experiments[186] found that while the non-equilibrium transition itself is not driven by heating, the hysteresis is due entirely to local temperature increases and the temperature dependence of the switching field. Moreover, from the timescales observed in the pulsed experiments, the heating is more extensive than just the magnetite channel material, affecting appreciable volumes of the MgO substrate proximate to the channel.

In the pulsed regime, when heating effects are minimized, we find that $V_{sw}$ does not approach zero as the length $L$ of the device approaches zero. This is a signature of contact resistances, with voltage drops taking place at the metal/magnetite interfaces. These contact effects, usually neglected or deliberately avoided in most materials characterization experiments, can provide valuable information about both the conduction mechanism in the bulk of the material and the mechanism of the field-driven breakdown. We performed a series of multi-terminal measurements to characterize the contact resistances and their dependence on the metal used for the injecting and collecting electrodes. **Figure 25** summarizes these results. We find that contact resistance is directly proportional to the magnetite channel resistance over the whole temperature range, both above and below $T_V$. That constant of proportionality depends on the particular contact metal, with an apparent trend of smaller contact resistances for higher work function metals.

From the perspective of conventional crystalline semiconductor materials, for example, this relationship between contact and bulk resistance is surprising. In a metal-semiconductor contact one would expect an additional temperature dependence of the contact resistance to arise due to thermal activation over the Schottky barrier at the interface. However, in highly disordered semiconductors such as organic polymers like polythiophene, a contact resistance proportional to the bulk resistance is observed.[188-190] In that case, transport in the bulk is dominated by hopping through a steeply energy dependent density of localized states, as disorder broadens out the valence band. Contact resistance in such a hopping conductor can originate from a competition between diffusion of an injected carrier away from the interface (proportional to the mobility and thus inversely proportional to the bulk resistivity) and attraction between that carrier and its image potential in the metal.[191]

The situation is different in the case of magnetite, ideally a crystalline material lacking this sort of severe disorder. However, conduction in magnetite is thought to proceed with a significant hopping contribution, due to strong correlations and the polaronic character of the carriers.[192, 193] These contact resistance measurements show consistency with the idea that the hopping conduction has the same effect on contact resistance even when the hopping results from correlations rather than disorder.

Looking at the contact resistance at the point of the non-equilibrium transition, we find that the contact resistance and channel resistance both drop dramatically as the bias exceeds $V_{sw}$. While it is difficult to avoid self-heating in the "on" state, the decrease in both injecting and collecting contact resistances is consistent with a Landau-Zener-type breakdown of the correlated state,[194, 195] as opposed to other resistive switching phenomena observed in transition metal oxides.[196, 197] This breakdown mechanism is credible since in the presence of the dc electric field (on the order of $10^7$ V/m), a charge carrier crossing one unit cell (approximately 0.8 nm) acquires a kinetic energy comparable to $k_B T_V$.

The situation must be a bit more complicated, however. When breakdown experiments are repeated in a given device many times at a fixed temperature, it is clear that there is a distribution of $V_{sw}$ values, and this distribution evolves with temperature and, surprisingly, magnetic field. If this were entirely a simple breakdown process, then increasing temperature would tend to broaden this distribution due to thermal fluctuations in carrier energy. Instead, the converse is observed, with the distribution of $V_{sw}$ increasing in width as temperature is decreased. Similar phenomenology has been observed in the kinetics of field-driven magnetization reversal in magnetic nanostructures;[198-200] in that case, a disorder landscape pins the propagation of the domain wall (boundary between the old phase and that favored by the driving field). This suggests that analogous disorder may be relevant in the stability of the Verwey state under a perturbing electric field. The role of defects in Landau-Zener-like breakdown of correlated states has already been considered by Sugimoto, Onoda, and Nagaosa.[195] However, a general treatment of field-driven destabilization of a state, given by Garg,[201] finds that the width of the distribution of switching field can scale with the magnitude of the switching field itself, raising the possibility that the increasing width of the $V_{sw}$ distribution as $T \to 0$ in our experiments could be intrinsic. A detailed theoretical investigation and further experimental probes of the breakdown process in this material would be necessary to clarify matters.

The case studies of $VO_2$ and $Fe_3O_4$ demonstrate the power of applying nanostructure techniques to strongly correlated materials. In the former material, the lack of gate response in nanowire-based, ionic liquid-based transistor structures reveals that it is unwise to consider the insulating state to be a simple semiconductor, despite the activated temperature dependence of the electrical conductance. Hydrogen doping via catalytic spillover in other $VO_2$ nanostructures does show that it is possible, in nanomaterial form, to manipulate the delicate balance between the competing correlated metallic and insulating phases. In the latter material, nanostructure-based experiments have revealed a non-equilibrium transition between the correlated ground state and a more conducting state. Here the balance between competing phases has been tipped by the application of a large electric field, made readily accessible through the small inter-electrode spacing in the devices. Measurements in similar structures have also shown that contact resistances, usually evaded in most macroscale studies of electronic properties, can reveal much about the conduction process at work in the material near equilibrium, and the nature of the observed non-equilibrium transition.

The prospects for the future are bright. Strongly correlated materials have been comparatively neglected in industrial applications and nanoelectronics in particular, even though correlations can lead to desirable properties and phase transitions. Integration with electronics has been limited in part by materials compatibility issues, including chemical stability throughout lithographic patterning, deposition, and etching processes. Vanadium dioxide and magnetite happen to be comparatively robust, while retaining electronic properties that could be of great use. Fortunately, processing of "challenging" materials has been advancing of late, driven by the rise of graphene, a material that, because of its sensitivity to reactive oxygen species and adsorbates, requires cleaning procedures different from those traditionally used with conventional semiconductors. The development of general methods of exfoliating layered materials,[202] including layered dichalcogenides like those discussed in the previous section, opens many opportunities for nanostructure studies of these systems and integration of these materials with technology on large scales.

**6. Looking Forward**

The materials that have been discussed here are tied together by a unifying thread of underlying physics. Electron-electron interactions play a dominant role in determining electronic and thermodynamic properties in these diverse systems. The result is a rich panoply of competing phases with distinct properties, separated by phase transitions that may be classical (controlled by thermal fluctuations) or quantum in nature (existing at $T = 0$, with quantum fluctuations persisting to higher temperatures). Through control of material parameters such as carrier density, pressure, and magnetic field, it is possible to tune the relative strengths of the interactions between constituent degrees of freedom. This alters the balance between competing phases, revealing the existence of new phases (*e.g.*, superconductivity) and elucidating the underlying physics at work. Nanoscale device-based examinations of these materials enable experiments to probe the stability of various phases; perturbing electric fields may be used to modulate carrier densities at fixed disorder in gating experiments, while applied electric fields can drive the material out of equilibrium with modest applied voltages; chemical doping may be studied on a microscopic level.

Because of the competition between phases with dramatically different properties, strongly correlated materials have much to offer as systems of potential technological interest. Systems poised at the transition between phases are exquisitely sensitive to small stimuli. Tipping the balance through some perturbation essentially leverages the thermodynamics and cooperative energetics of the underlying degrees of freedom to produce a large response. This sensitivity suggests that strongly correlated materials are well positioned to be active materials in technologies. An example of this is the realization that correlated manganites, at the border between a paramagnetic insulating state and a ferromagnetic metallic state, are tremendously sensitive to magnetic field, exhibiting colossal magnetoresistance (CMR),[203-205] and might therefore be of interest in magnetic data storage technologies. While this particular application has not materialized commercially, the basic principle is sound.

Moreover, the timescales inherent in the correlated materials of interest tend to be very rapid. Purely electronic transitions driven by eV-scale electron-electron interactions can have intrinsic timescales as short as $\sim \hbar/(1\text{eV}) \sim 1$ fs. For example, the electronic transition in $VO_2$ has been observed to take place on femtosecond timescales when driven by ultrafast optical pulses.[206] Even when strong electron-vibrational couplings are important, the intrinsic transition timescale (picoseconds) is fast enough to be of technological interest. Thus strongly correlated materials, if properly engineered and interfaced with other systems, are exciting and relatively unexplored possibilities when considering the technological need for active materials with high intrinsic speeds and high sensitivity to stimuli such as temperature, electric and magnetic fields, and light.

Our improved understanding of these materials from the theoretical and experimental points of view is well timed, arising concurrently with the development of new computational techniques for the predictive understanding of complex materials. There is a growing recognition that we may be on the verge of an enormous boon in the ability to model materials from the level of quantum chemistry at the atomic scale to macroscale treatments of bulk properties. This is precisely the justification for the recent call by the US Office of Science and Technology Policy for the launching of a Materials Genome Initiative[1], with the intent to do for predictive materials design and engineering what the Human Genome Project has done for genetic research. The eventual goal is "materials by design": A feedback loop (see **Fig. 26**) including theoretical understanding, computational modeling, materials synthesis and discovery, and materials characterization across energy and timescales, would lead to the formulation of design principles and the creation of new materials systems with specific technical goals in mind. Specific examples where strongly correlated materials may figure prominently are high temperature superconductors and hard magnets without rare earth elements. In the former case, the potential technological impact of higher temperature superconductors is enormous, provided that such materials can be made with favorable combinations of transition temperature, critical current, critical magnetic field, and mechanical properties. In the latter case, our improved knowledge of local moment formation and exchange physics suggests paths forward in the quest for novel magnets, and compounds such as $Fe_{1/4}TaS_2$ discussed above are promising starting points.

Many challenges remain on the path toward intellectual understanding of strongly correlated materials and their incorporation into technologies, however. As a matter of principle, we are only beginning to address the question: are there unifying principles that underlie a variety of strongly correlated systems? Quantum criticality has shown considerable promise to be one principle of this kind. As such, considerable effort exists to further understand and even classify quantum critical points. With the increasing list of new materials being developed, new properties being uncovered by experiments, and new developments in theory, there is room for optimism that new unifying principles will be discovered and understood. From the computational perspective, it is still extraordinarily challenging to perform quantum chemical calculations of strongly correlated materials that fully capture the importance of the correlation physics, as mentioned in our discussion of $VO_2$. The tradeoff between computational accuracy and scalability remains extremely stark. Nevertheless, the continuing advances in algorithm development and powerful many-body techniques, such as the cluster dynamical mean-field theory (cluster-DMFT)[207] and the recent progress in the self-consistent GW approach,[208] for example, are very promising. In particular, the combination of the dynamical mean-field theory with the input from *ab initio* electronic structure within the so-called LDA+DMFT framework[72-74] gained popularity and has seen a number of successful applications to correlated electron compounds recently. These exciting theoretical developments, while not a panacea, make closer the prospect of the accurate prediction of electronic properties in strongly correlated materials.

Integration of strongly correlated materials with existing technologies is also not trivial. Even in systems known for a long time, such as the superconducting cuprates, it has taken years to develop scalable manufacturing techniques that can accommodate the complex stoichiometry, challenging chemical stability, and mechanical fragility of the material. As should be clear from our discussion, the balance between competing phases in this class of materials is readily tipped by composition, strain, disorder, temperature, and external electric and magnetic fields. As such, these systems are considerably more challenging to grow and integrate than, e.g., conventional compound semiconductors. However, there is little doubt that sufficiently desirable properties can be a powerful motivation for the development of integration techniques. The successful technological deployments of magnetic tunnel junctions in hard drive read heads, piezoelectric and electro-optic lithium niobate in optical modulators, ternary phase change compounds in recordable optical drives, and nonlinear optical crystals in devices as trivial as laser pointers all speak to the adaptability of designers to accommodate challenging materials when sufficiently motivated. The prospects for strongly correlated materials are bright, both in terms of intellectual understanding and technological applications. True progress is enabled by the coherent, close coupling of theory and experiment, from model Hamiltonians and computational techniques to materials growth, discovery, and detailed characterization. We authors have been fortunate that Rice University fosters such an environment, and look forward to greater accomplishments in these materials in the years to come.

**Acknowledgements**

This work has been supported in part by PECASE (EM), US DOE grant DE-FG02-6ER46337 (DN), and NSF Grant No. DMR-1006985 (QS). The authors acknowledge recent collaborations and discussions with R. J. Cava, H. W. Zandbergen and K. Wagner (EM), P. Coleman and G. Kotliar (AHN), A. A. Fursina, S. Lee, J. Wei, H. Ji, I. V. Shvets and R. G. S. Sofin (DN), and R. Yu, J. Pixley, J. Wu, E. Nica, P. Goswami, S. Kirchner, S. Yamamoto, L.

Zhu, E. Abrahams, J. Dai, K. Ingersent, J.-X. Zhu, F. Steglich, S. Paschen, J. Custers, S. Friedemann, P. Gegenwart, O. Stockert, and S. Wirth (QS).

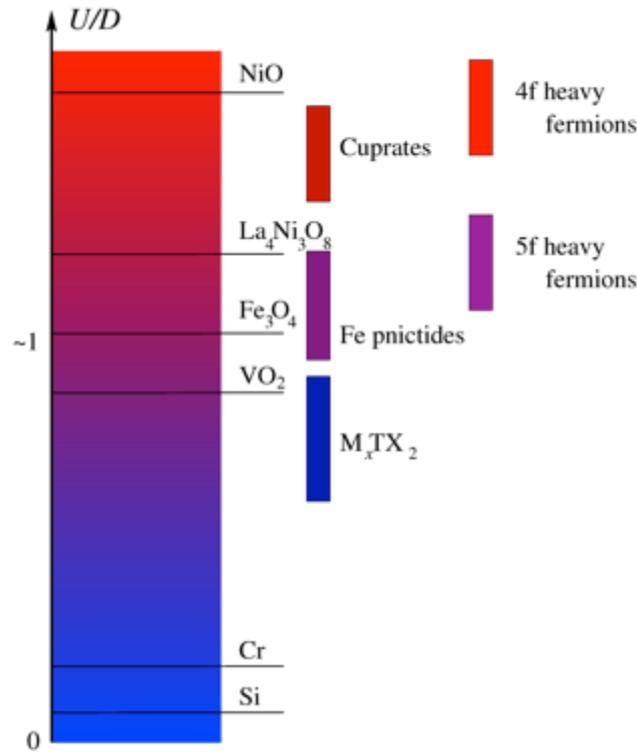

**Figure 1.** The chart of various correlated electron materials discussed in this article, plotted as a function of the on-site Hubbard interaction strength $U$ relative to the electron bandwidth $D$. The value of the $U/D$ varies between different compounds in the series (*e.g.*, in the iron pnictides or dichalcogenides $M_xTX_2$) and the displayed values should be understood as semi-qualitative. The color code serves as a rough guide of the interaction strength, increasing from blue (non-interacting) to red (very strongly interacting).

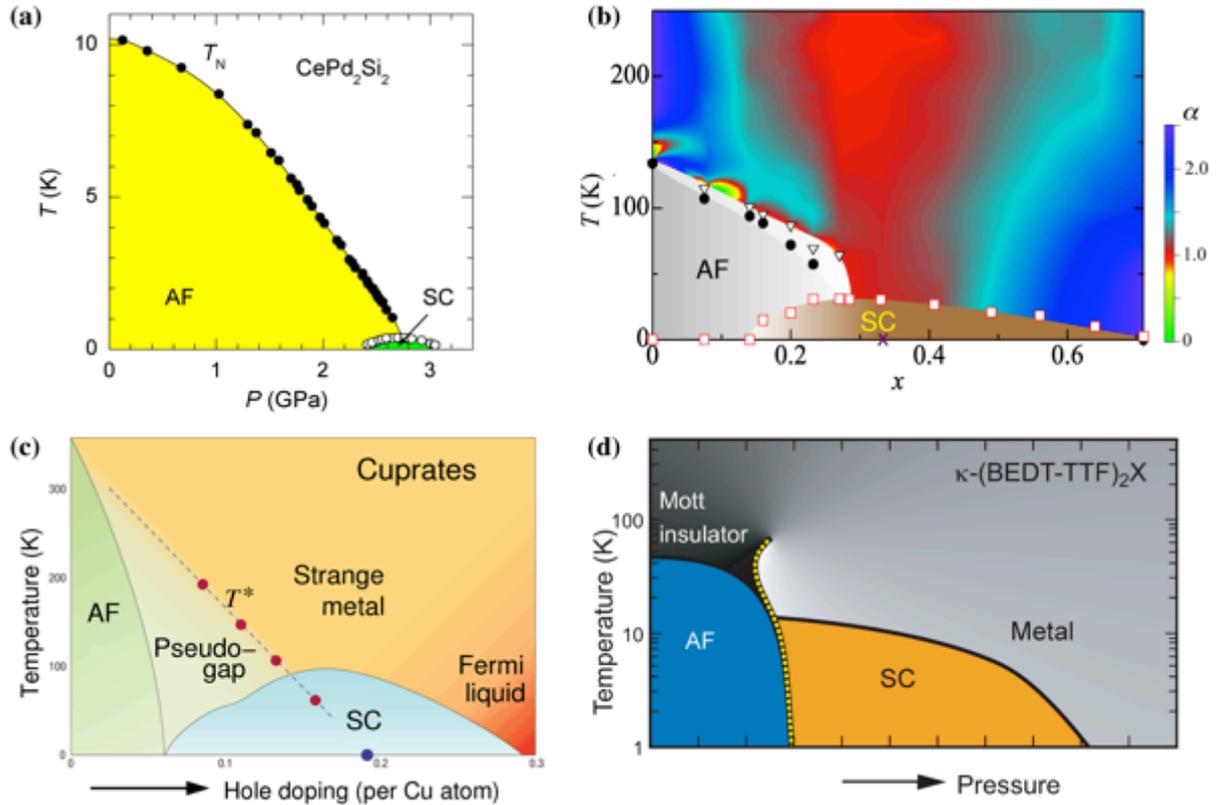

**Figure 2.** Antiferromagnetic order (AF) and superconductivity (SC) in strongly correlated systems. Shown are the phases of (a) the heavy fermion compound $CePd_2Si_2$ as a function of temperature ($T$) and pressure ($p$) (reproduced with permission from Ref. [2]. Copyright 1998, Nature Publishing Group.). $T_N$ marks the Néel temperature; (b) the iron pnictide $BaFe_2(As_{1-x}P_x)_2$ as a function of $T$ and the concentration $x$ of the isoelectronic P-doping for As (adapted from Ref. [3]). The color shading (legend on the right) represents the resistivity exponent $\alpha$ appearing in the temperature dependence of the electrical resistivity, $\rho = \rho_0 + AT^\alpha$. On the horizontal axis, the cross denotes the QCP at $x_c \approx 0.33$; (c) the copper oxides, illustrated as a function of $T$ and carrier doping concentration (adapted from Ref. [4]). "Strange metal" denotes a non-Fermi liquid regime appearing near the optimal doping for superconductivity. "Pseudogap" and "Fermi liquid" respectively describe the regimes in the normal state at under- and over-dopings; and (d) the organic compound, $\kappa$-$(BEDT-TTF)_2X$, as a function of pressure (adapted from Ref. [5]).

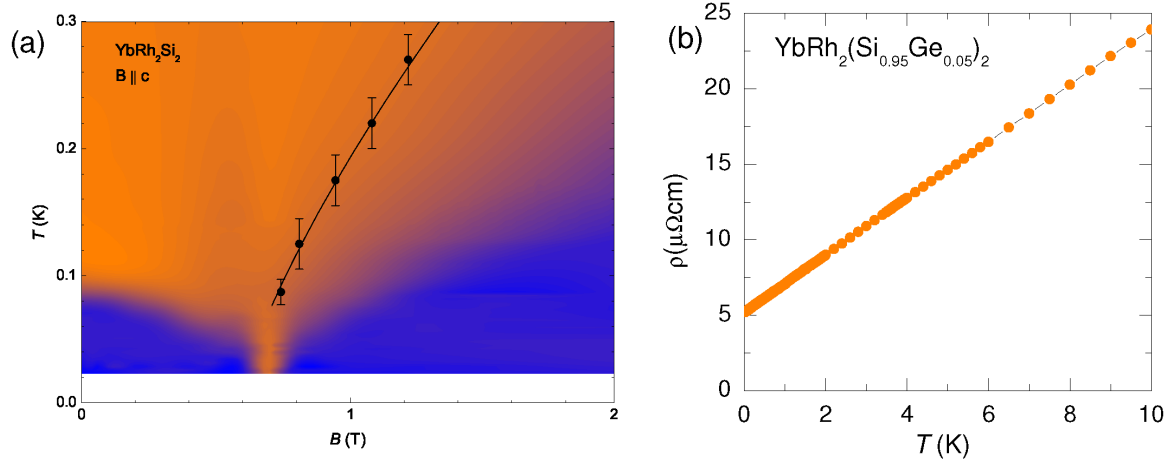

**Figure 3.** Quantum critical behavior in heavy fermion metals. (a) Temperature-magnetic field (T -B) phase diagram of the heavy fermion compound YbRh$_2$Si$_2$, showing a QCP at the critical magnetic field $B_c$=0.7 T. The temperature dependence of the electrical resistivity has the Fermi-liquid $T^2$ form in the blue regions, corresponding to $T < T_N$ at $B < B_c$ and $T < T_{FL}$ at $B > B_c$. It becomes linear in the orange regime, illustrating the non-Fermi liquid behavior in the quantum critical regime. The solid line is described in the text; (b) The electrical resistivity in the quantum critical regime of the slightly Ge-doped YbRh$_2$Si$_2$ shows a $T$-linear dependence over a wide temperature range of about three decades. It illustrates quantum criticality as a robust mechanism for non-Fermi liquid electronic excitations. Data from Ref. [19].

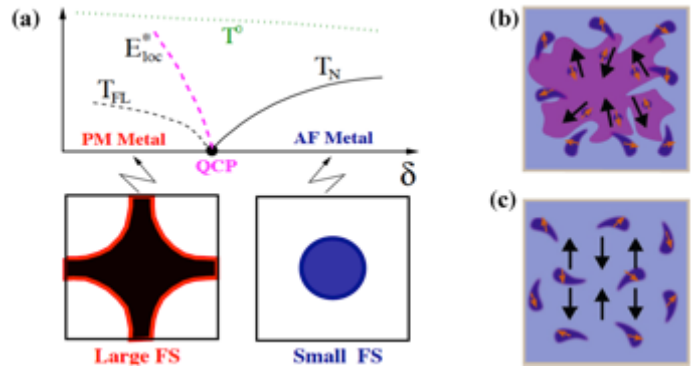

**Figure 4.** (a) Schematic phase diagram for local quantum criticality. The $E^*_{loc}$ energy scale defines two regimes. In one, the system goes towards a Kondo-screened paramagnetic metal ground state, a heavy-fermion state whose effective Fermi energy is marked by $T_{FL}$. In the other, the system flows towards a Kondo-destroyed antiferromagnetic metal state, which has a Néel temperature $T_N$. The Kondo-screened state has a large Fermi surface (*f*-electrons itinerant, lower left inset), while the Kondo-destroyed one has a small Fermi surface (*f*-electrons localized, lower right inset). Here δ is the ratio of the RKKY to Kondo interactions. $T^0$ marks the scale for the initial onset of Kondo screening; (b) The Kondo singlet ground state (the purple profile) between the localized moments (the thick black arrows) and the mobile conduction electrons (the orrange arrows); (c) Kondo breakdown in the antiferromagnetic phase. The local moments arrange into an antiferromagnetic order among themselves, and the Kondo singlet has been destroyed. Adapted from Ref. [26].

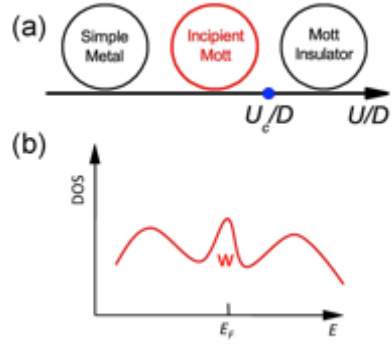

**Figure 5.** (a) Three regimes of electronic correlations; (b) The decomposition of the single electron spectral function, shown here as a function of energy $E$, as the sum of the coherent and incoherent parts. $w$ denotes the spectral weight in the coherent part. Each of the peaks may contain additional fine structure because of the multi-orbital nature of the iron pnictides and chalcognides. Note that $w$ grows as $U/D$ is reduced.

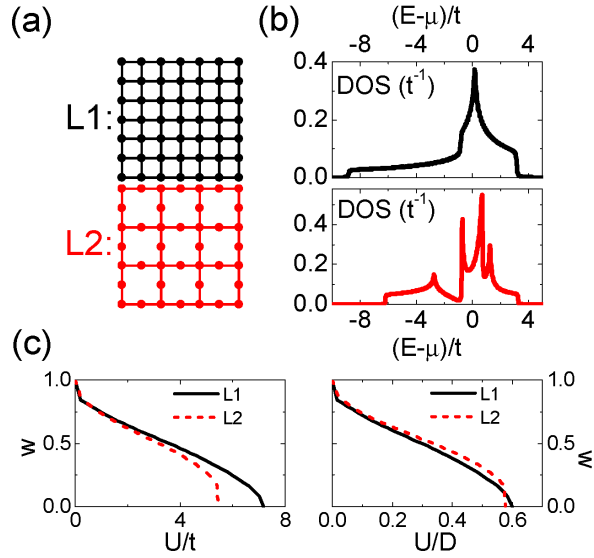

**Figure 6.** Kinetic energy reduction and Mott localization by ordered Fe vacancies. (a) The Fe square lattice and one with a $2 \times 2$ vacancy order; (b) The density of states of the Fe square lattice and its counterpart for the vacancy ordered lattice; (c) Left panel: the coherent weight of electronic excitations plotted against $U/t$ showing that the Mott threshold $U_c$ is reduced by the ordered vacancies, in a simplified model. Here, $t$ is the tight-binding hopping parameter between nearest-neighbor sites. Right panel: the same result, but plotted against $U/D$, where $D$ is the bandwidth for each case. The reduction of $U_c$ of the vacancy-ordered lattice tracks its reduction in $D$, showing that the enhanced tendency towards Mott localization originates from the kinetic energy reduction. Adapted from Ref. [43].

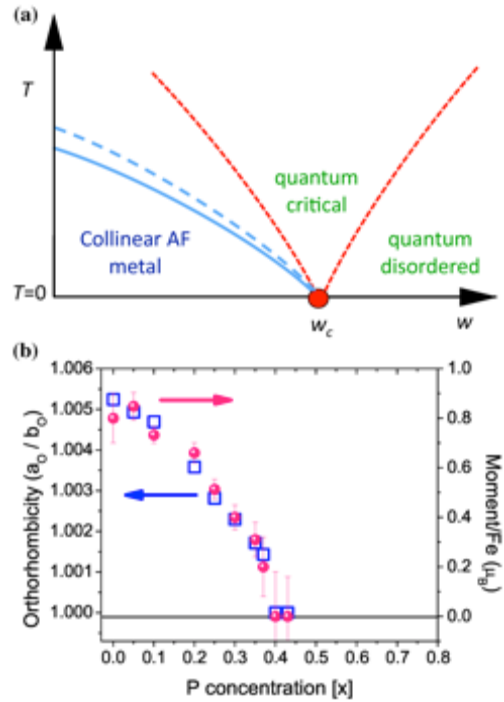

**Figure 7.** (a) Quantum critical point induced by the increasing *w* associated with an enhancement of the kinetic energy (With permission from Ref. [55]. Copyright 2009, National Academies Press.); (b) Continuous suppression of the collinear antiferromagnetic order and the tetragonal-to-orthorhombic structural distortion induced by P-substitution for As in Ce1111 compounds (With permission from Ref. [56]. Copyright 2010, American Physical Society).

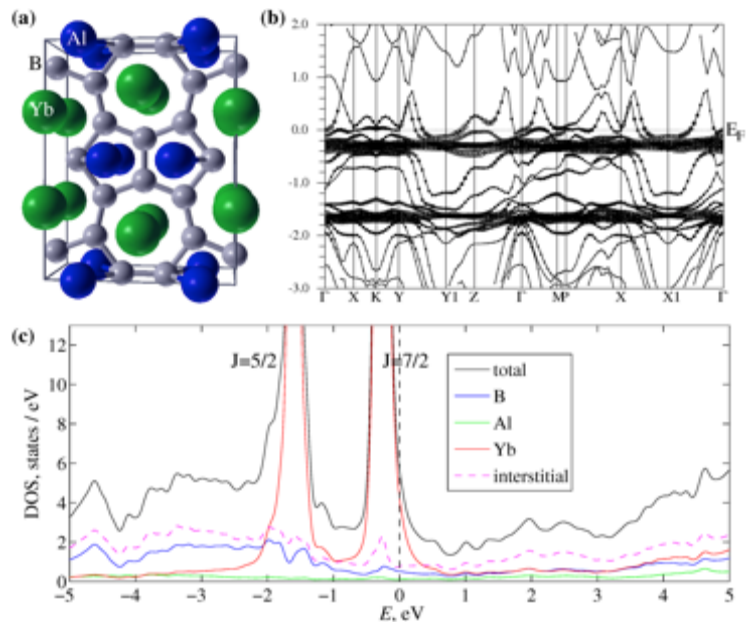

**Figure 8.** (a) Atomic structure (conventional unit cell) of β-YbAlB$_4$. (b) The calculated band structure along high-symmetry lines in the irreducible Brillouin zone. Shown are the so-called "fat bands", with the dot size proportional to the amount of Yb-*f* character in the given band. (c) The projected density of states, showing individual contributions of different atoms. The narrow Yb *f*-bands form two clear peaks below the Fermi level, corresponding to angular momentum *J* = 5/2 and 7/2, split by spin-orbit coupling.

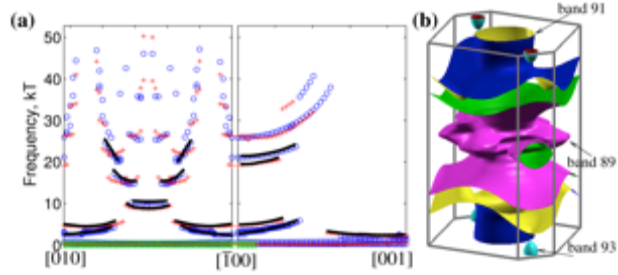

**Figure 9.** (a) The predicted de Haas-van Alphen frequencies for β-YbAlB$_4$, based on band structure calculation of the extremal orbits on the Fermi surface (circles). The black lines are guides to the eye for the least massive bands with effective mass $m^*/m_e < 10$. (b) The calculated Fermi surface of β-YbAlB$_4$, showing two large area sheets (bands 89 and 91) and a small pocket (band 93).

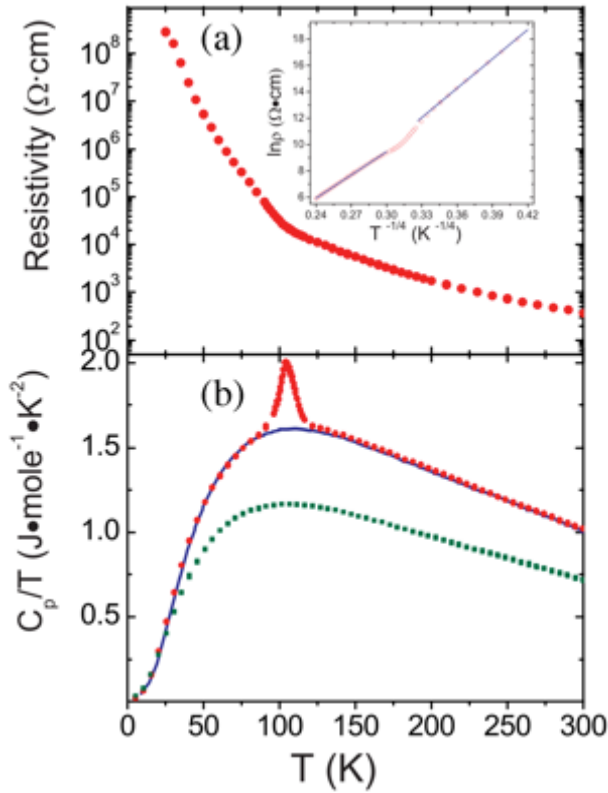

**Figure 10.** (a) Temperature dependence of resistivity (logarithmic scale). The inset shows ln(ρ) versus $T^{-1/4}$ with observed data (circles) and line fits (solid lines) according to Mott's variable range-hopping model. (b) Specific heat data of La$_4$Ni$_3$O$_8$ (circles) and La$_3$Ni$_2$O$_6$ (squares). The solid line represents the scaled specific heat of La$_3$Ni$_2$O$_6$. (Reproduced with permission from Ref. [81]. Copyright 2010, American Physical Society.)

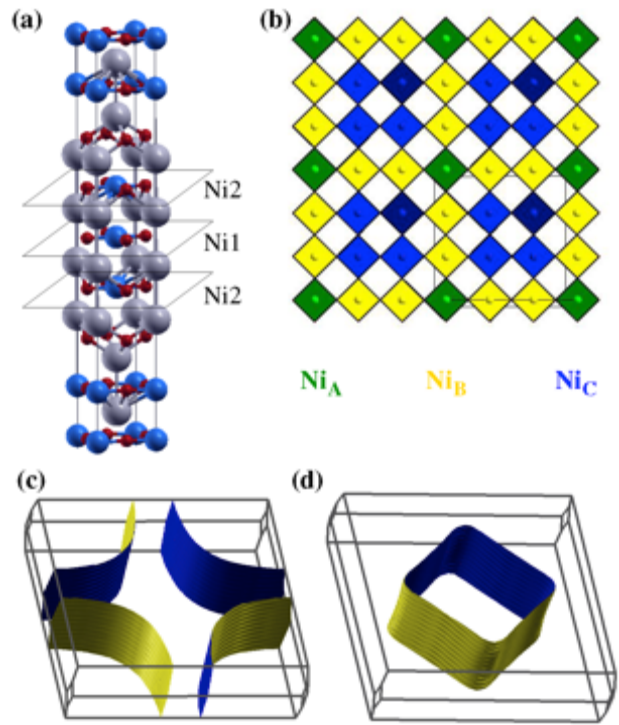

**Figure 11.** (a) Crystal structure of $La_4Ni_3O_8$, showing inequivalent layers $Ni(1)O_2$ and $Ni(2)O_2$. (b) Theoretically proposed magnetic structure of $La_4Ni_3O_8$, showing the in-plane structure of the $Ni(1)O_2$ layer (the $Ni(2)O_2$ layers order in a similar pattern) with three inequivalent colour-coded magnetic sublattices (A, B, and C). The thin black line denotes the magnetic supercell boundary with 9 times the volume of the paramagnetic unit cell. (c, d) The Fermi surface sheets from the GGA band theory calculations in the paramagnetic state

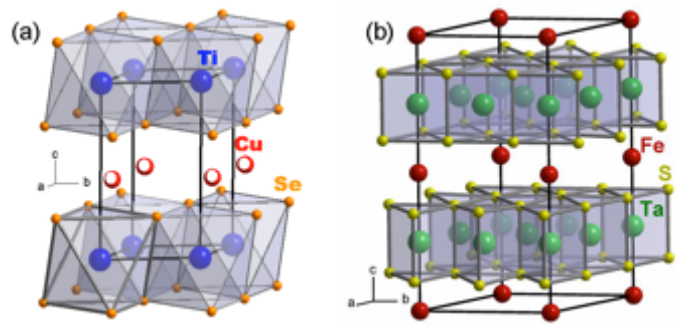

**Figure 12.** Crystal structure of (a) $TiSe_2$ and (b) $TaS_2$, with octahedral or trigonal prismatic coordination of the transition metal, respectively. The intercalant (red) occupies the Van der Waals positions in-between the $TX_2$ layers.

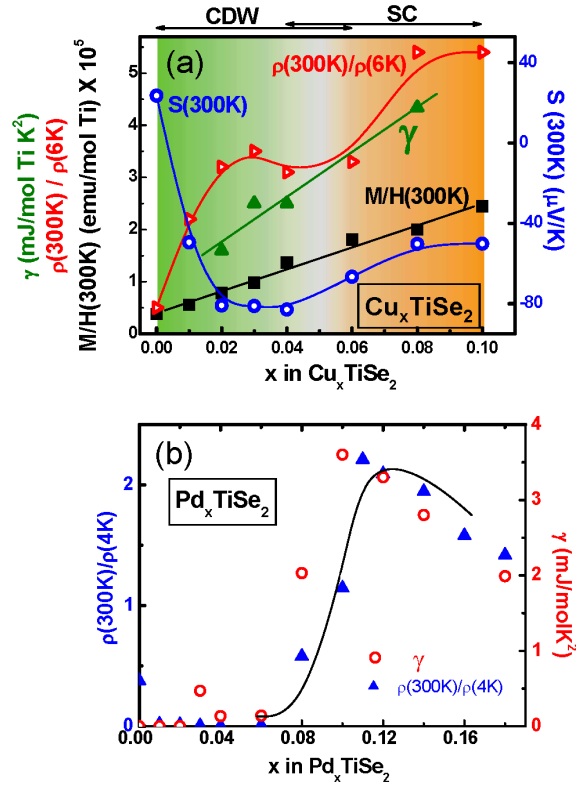

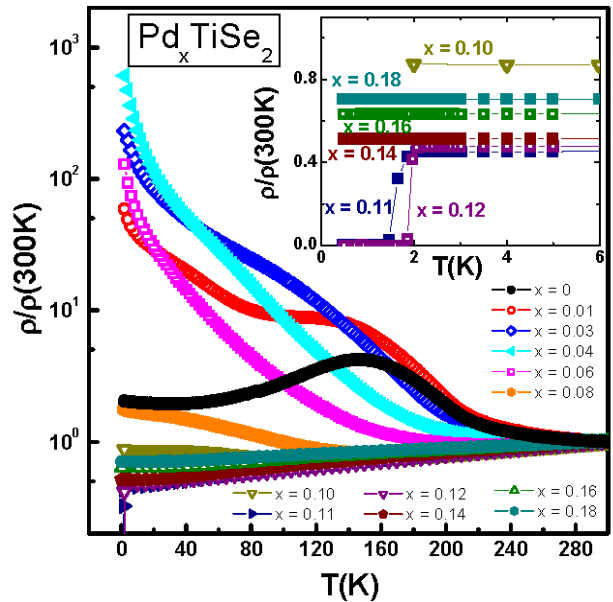

**Figure 13.** Temperature dependent (a) specific heat coefficient γ, scaled resistivity ρ(300K)/ρ(6K), room temperature susceptibility *M/H*(300K) and Seebeck coefficient *S* for Cu$_x$TiSe$_2$ and (b) specific heat coefficient γ and scaled resistivity ρ(300K)/ρ(6K) for Pd$_x$TiSe$_2$.

**Figure 14**. Temperature-dependent resistivity data (scaled at 300 K) for Pd$_x$TiSe$_2$ on a semilog plot; inset: scaled low-temperature ρ(T) data for 0.10 ≤ *x* ≤ 0.18, showing the superconducting transitions for *x* = 0.11 and 0.12.

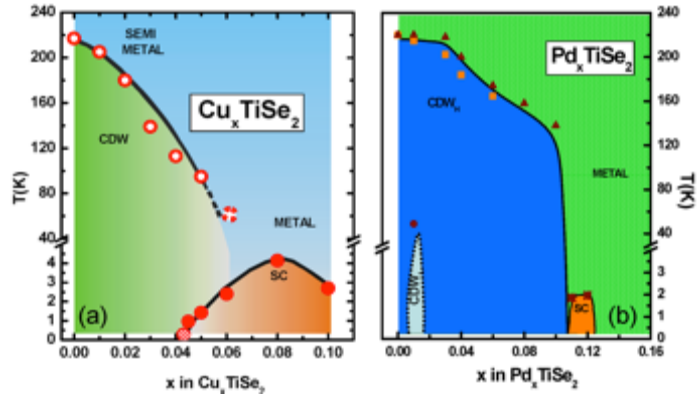

**Figure 15.** The *T-x* electronic phase diagram for (a) $Cu_xTiSe_2$ and (b) $Pd_xTiSe_2$.

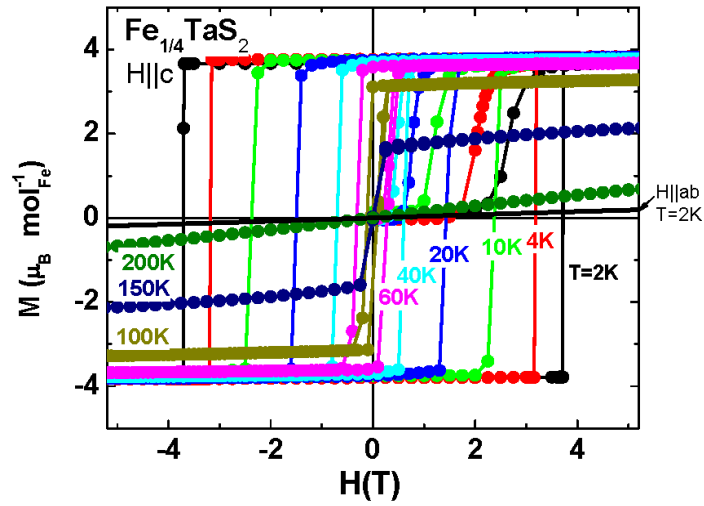

**Figure 16.** $Fe_{1/4}TaS_2$ *H* ∥ *c* (symbols) *M(H)* loops for temperatures between 2 and 200 K. Also shown is the *H* ∥ *ab* *M(H)* *T* = 2 K curve (solid line), illustrating the strong magnetic anisotropy.

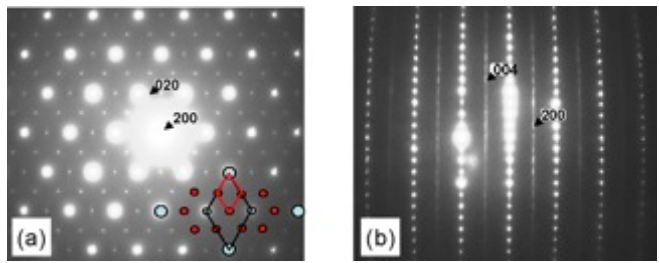

**Figure 17.** Electron diffraction patterns of the $Fe_{1/4}TaS_2$ reciprocal lattice (a) along the [001] and (b) perpendicular to the [001] direction. The basic trigonal structure and the 2*a* superstructure reflections seen in (a), with the supercell outlined and the unit cell indicated; in (b) strong streaking of the superreflections along *c** can be seen.

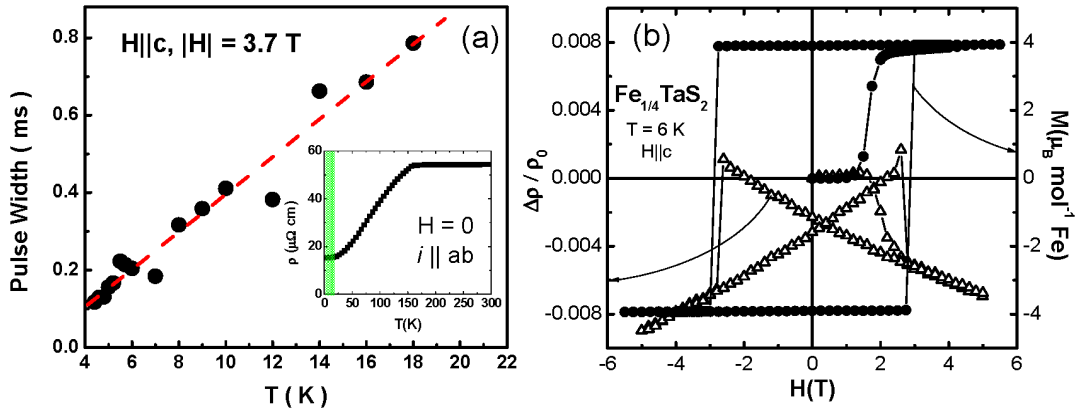

**Figure 18.** (a) The pulse width in $Fe_{1/4}TaS_2$ varies linearly with temperature at low *T*, where the resistivity (inset) is nearly *T*-independent. (b) The *T* = 6 K $H \parallel c$ magnetoresistance (open symbols) displays a sharp drop around the switching field $H_s$ where the magnetization (full symbols) rapidly switches direction.

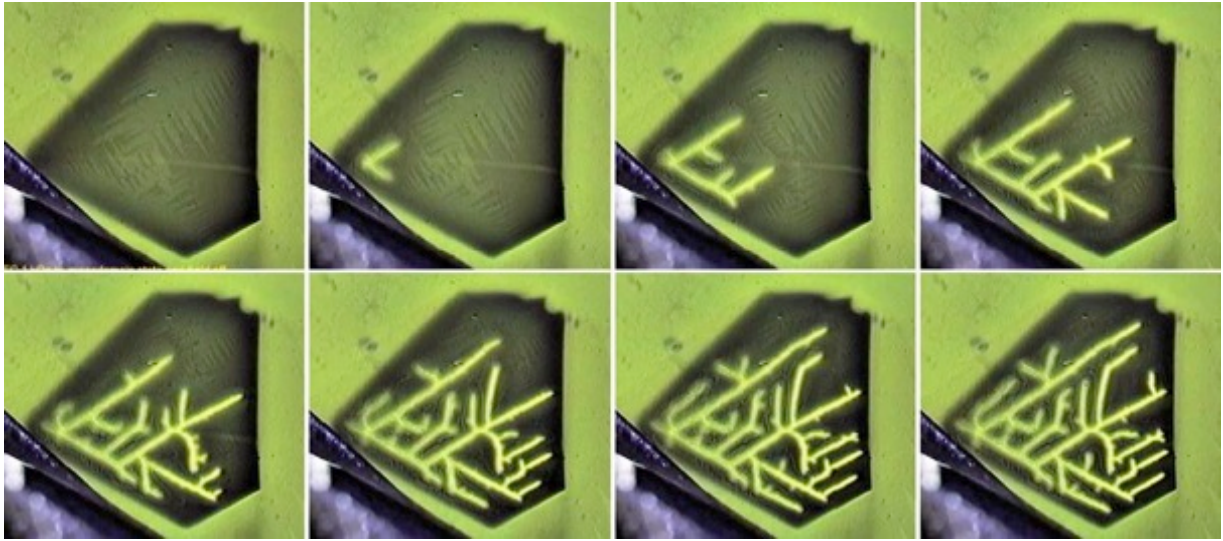

**Figure 19.** T = 5 K field-cooled monodomain in $Fe_{1/4}TaS_2$ (top left), and dendritic domain nucleation as an increasing magnetic field is applied. Left to right: top row *H* = 0, 500, 600, 700 Oe; bottom row *H* = 850, 1000, 1200, 1500 Oe.

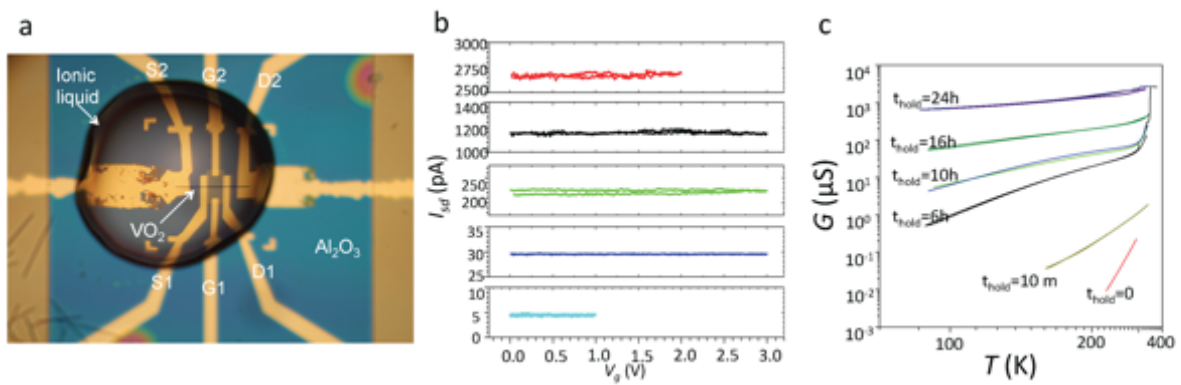

**Figure 20.** Ionic liquid gating of $VO_2$ nanowires. (a) Optical micrograph of the experimental arrangement. Source and drain electrodes (S1, D1, S2, D2) are indicated, as are gate

electrodes (G1, G2). A protective aluminum oxide layer (blue) has been deposited to limit the contact between the ionic liquid and the source/drain electrodes; a hole has been patterned in the aluminum oxide to allow contact between the ionic liquid and the nanowire. (b) Source-drain current measured via lock-in with 1 mV AC (13 Hz) source-drain bias applied, as a function of gate voltage (swept at 2 V/min) when the ionic liquid is free from water contamination. No detectable gate response in the source-drain conduction is seen. (c) Source-drain conductance as a function of temperature (ticks on a reciprocal scale) following a sustained +1 V gate bias application in the presence of a water-contaminated ionic liquid in the $VO_2$ metallic state (383 K) for a duration $t_{hold}$. We ascribe the persistent change in conductance, absent when the ionic liquid is dry, to an electrochemical doping process.

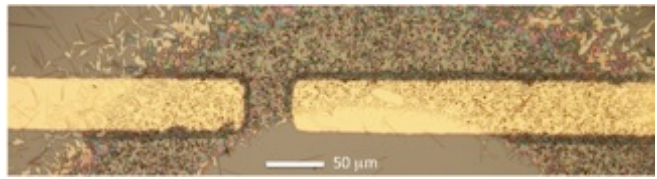

**Figure 21.** Spillover hydrogenation of $VO_2$ crystals. Room temperature optical micrograph showing crystals on oxidized Si substrate, with subsequently deposited Au pads, after exposure to hydrogen at 180°C for 20 minutes. The dark "halo" around the pads is composed of $VO_2$ crystals that have been hydrogenated into a stabilized metallic state.

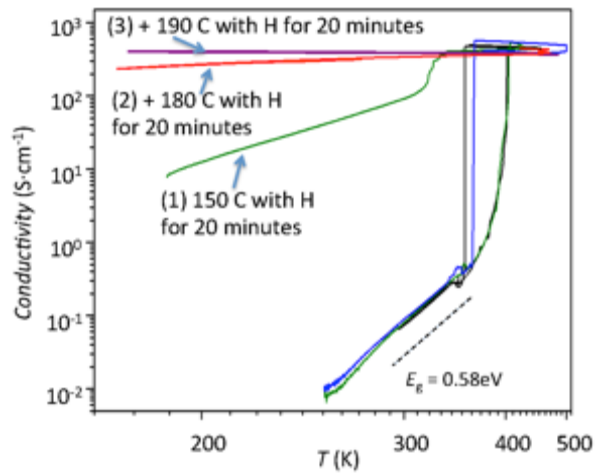

**Figure 22.** Conductivity as a function of temperature (ticks with reciprocal $T$ spacing) for a suspended $VO_2$ nanobeam with Ti/Au contacts under various hydrogenation conditions. The blue curve show the MIT and activated conduction in the insulating state consistent with an energy gap of ~ 0.6 eV prior to any hydrogen exposure. Exposure to hydrogen at moderate temperatures (> 150°C for tens of minutes) increases the conductivity and decreases the effective energy gap in the insulating state. After more than an hour of hydrogenation, the MIT has vanished and the device is metallic over the whole temperature range (purple curve at top). After baking at 250°C for 20 minutes in air, the device returns to its original electronic conduction properties (black curve).

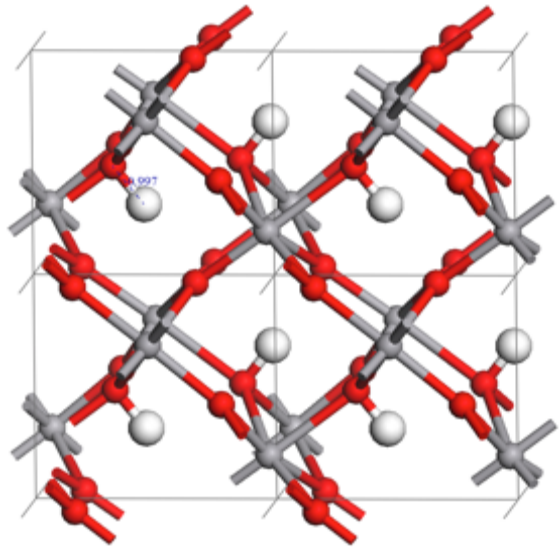

**Figure 23.** Theoretically optimized (within DFT-GGA[165]) atomic structure of the pseudo-rutile phase of $H_2V_4O_8$ (50 mol% H per $VO_2$). Oxygen atoms are shown in red, vanadium in grey and hydrogen in silver. The H..O bond length is very short, at 0.997 Å.

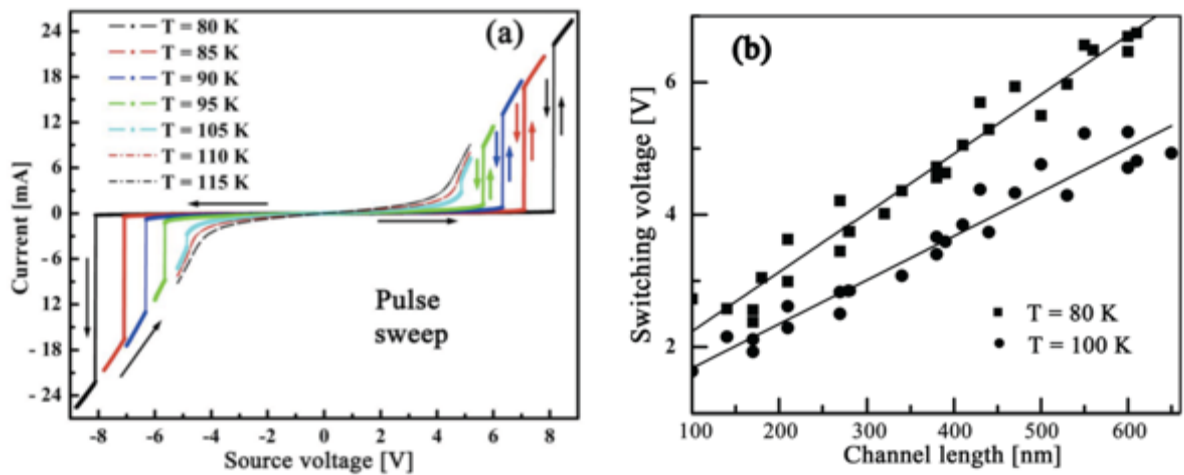

**Figure 24.** Non-equilibrium transition in magnetite nanodevices. (a) Current as a function of voltage for a series of temperatures, with a source-drain electrode spacing of approximately 500 nm, acquired using voltage pulses of 500 microsecond duration separated by 100 milliseconds. (b) Switching voltage between the high- and low-resistance states for a series of devices of different channel lengths. The linear scaling of $V_{sw}$ with channel length indicates that the non-equilibrium transition is driven by electric field, while the nonzero intercept as $L \rightarrow 0$ indicates that contact resistance is non-negligible. From A. A. Fursina's doctoral thesis [187].

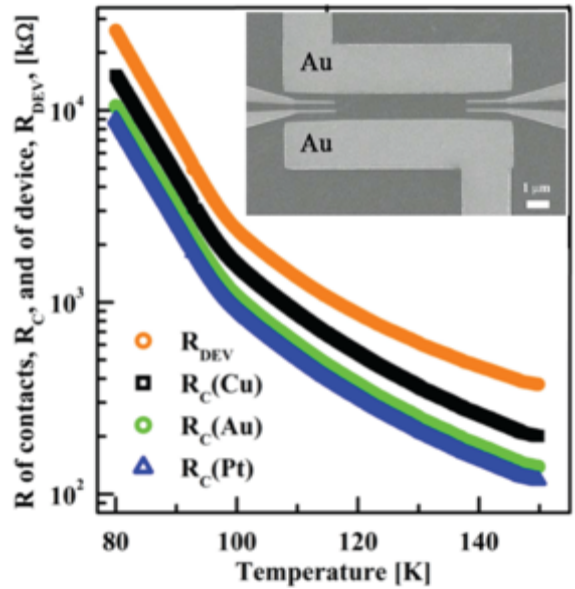

**Figure 25.** Contact resistance and bulk resistance of magnetite device (inset) as a function of temperature, for different contact metals. The fact that the contact and bulk resistances are directly proportional over the whole temperature range results from the hopping character of conduction in the magnetite.

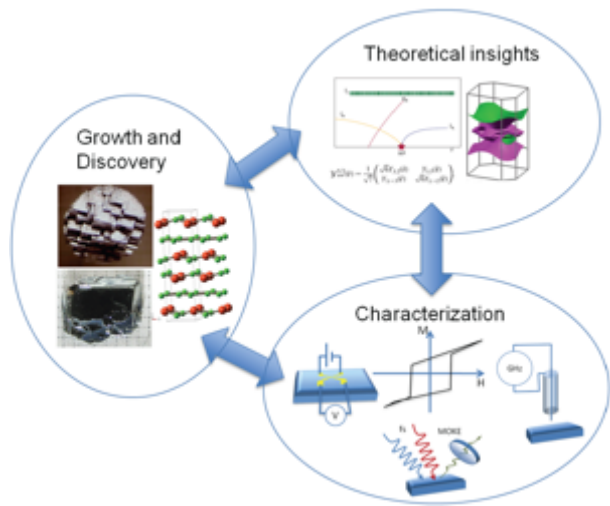

**Figure 26.** The feedback cycle required for progress toward materials by design. Theoretical insights, materials characterization, and materials growth and discovery all reinforce one another. The ideal final output of such a cycle would be novel materials with desired properties and design principles for the development of new material systems.

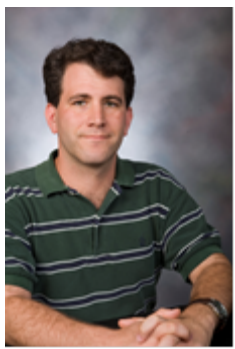
*Douglas Natelson is a professor of Physics and Astronomy and in Electrical and Computer Engineering at Rice University, where he arrived in 2000 after a doctorate at Stanford University and a postdoc at Bell Labs. His research focuses on experimental condensed matter physics at the nanoscale, including the properties of atomic- and molecular junctions and nanostructures made from strongly correlated materials. He has published over 80 reviewed papers, and blogs about nanoscale science at nanoscale.blogspot.com.*

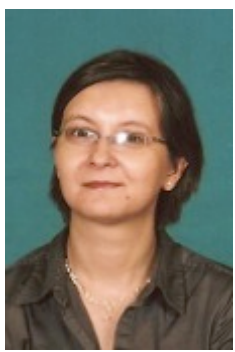
*Emilia Morosan is an assistant professor of Physics and Astronomy and Chemistry at Rice University. She has been at Rice since 2007, following doctoral work at Iowa State University and a postdoc at Princeton University. Her research in condensed matter experiment is focused on magnetism, superconductivity and quantum criticality. Prof. Morosan's research has been recognized with the 2009 Presidential Early Career Award for Scientists and Engineers (PECASE), and has authored over 50 reviewed publications.*

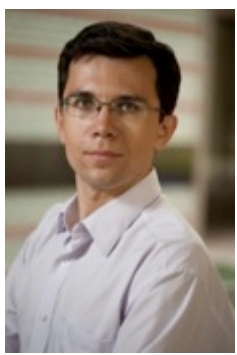
*Andriy Nevidomskyy (Ph.D. Cambridge University, 2005) is an assistant professor of Physics and Astronomy at Rice University, where he has been since 2010, after holding postdoctoral research fellowship at Université de Sherbrooke, Canada and at Rutgers University, New Jersey. His research is centered on theoretical understanding of strongly correlated electron systems, including magnets, unconventional superconductor and heavy fermion materials, for which Nevidomskyy uses a combination of analytical tools and first principles electronic band structure calculations.*

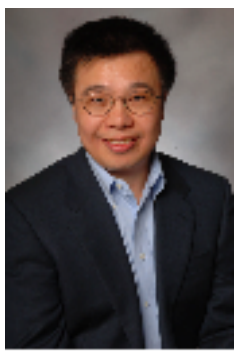
*Qimiao Si is the Harry C. and Olga K. Wiess Professor of Physics at Rice University. He obtained a B.S. (1986) degree from University of Science and Technology of China and a Ph.D. (1991) degree from the University of Chicago, and is known for theoretical work on quantum phase transitons and on unconventional and high temperature superconductivity. He is a Fellow of IOP (UK), APS, AAAS, and received a Sloan Fellowship and a Humboldt Research Award.*

**The table of contents entry** should be fifty to sixty words long, written in the present tense, and refer to the chosen figure.

**TOC entry:** The relative importance of electron-electron interactions, $U$, compared with the kinetic energy in the form of the bandwidth, $D$, delineates between weakly and strongly correlated materials. We discuss several types of strongly correlated materials, higlighting their rich physics and diverse properties. Our improved understanding of these systems opens the exciting possibility of controlling and applying their fascinating phases.

*Keyword (see list): superconductivity, magnetism, metal-insulator transition, strong correlations, electron-electron interactions*

Emilia Morosan, Douglas Natelson*, Andriy H. Nevidomskyy and Qimiao Si

*Strongly Correlated Materials*

*ToC figure ((55 mm broad, 50 mm high, or 110 mm broad, 20 mm high))*

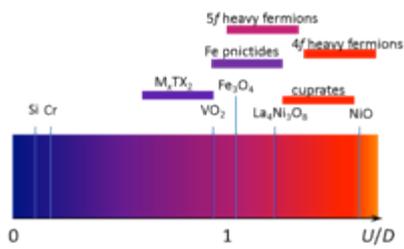